\documentclass[journal]{IEEEtran}
\IEEEoverridecommandlockouts                              
\def\JNT#1{{#1}}

\def\jt#1{{#1}}
\def\jnt#1{{#1}}

\usepackage{amssymb,amsmath}
\usepackage{amsfonts}
\usepackage[english]{babel}
\usepackage[latin1]{inputenc}
\usepackage{fancyhdr}
\usepackage{yfonts}
\usepackage{mathrsfs}
\usepackage[dvips]{graphicx}
\usepackage[rflt]{floatflt}
\usepackage{cite}

\usepackage{algorithm}
\floatname{algorithm}{Alg.}
\usepackage{algorithmic}
\usepackage{psfrag}

\providecommand{\com[2]}{\begin{tt}[#1: #2]\end{tt}}
\usepackage{color}

\newcommand{\modif}[1]{{#1}}
\newcommand{\modifnj}[1]{{#1}}
\newcommand{\modrev}[1]{{#1}}

\newtheorem{prop}{Proposition}
\newtheorem{thm}{Theorem}

\newtheorem{cor}{Corollary}
\newtheorem{conj}{Conjecture}
\newtheorem{lem}{Lemma}

\providecommand{\prt}[1]{\left( #1 \right)}

\providecommand{\abs[1]}{\left|#1\right|}
\providecommand{\norm[1]}{\| #1 \|}%

\providecommand{\ceil[1]}{\lceil #1  \rceil}
\providecommand{\1}{\textbf{1}} \providecommand{\quotes[1]}{``#1"}
\providecommand{\scalprod[2]}{\left<#1,#2\right>}%

\usepackage{epsfig}

\title{On Krause's multi-agent consensus model with state-dependent
connectivity (Extended version)}

\author{Vincent D.
Blondel, Julien M. Hendrickx and John N. Tsitsiklis
\thanks{This research was supported by the National Science Foundation under grant ECCS-0701623, by
the Concerted Research Action (ARC) \quotes{Large Graphs and
Networks} of the French Community of Belgium and by the Belgian
Programme on Interuniversity Attraction Poles initiated by the
Belgian Federal Science Policy Office. The scientific
responsibility rests with its authors. Julien Hendrickx holds
postdoctoral fellowships from the F.R.S.-FNRS (Belgian Fund for
Scientific Research) and the B.A.E.F. (Belgian American Education
Foundation); a part of this research was conducted when he was
with the Universit\'e catholique de Louvain. \hfil\break V.\ D.\
Blondel is with Department of Mathematical Engineering,
Universit\'e catholique de Louvain, Avenue Georges Lemaitre 4,
B-1348 Louvain-la-Neuve, Belgium; {\tt\small
vincent.blondel@uclouvain.be}.  J.\ M.\ Hendrickx and J.~N.\
Tsitsiklis are with the Laboratory for Information and Decision
Systems, Massachusetts Institute of Technology, Cambridge, MA
02139, USA; {\tt\small jm\_hend@mit.edu, jnt@mit.edu}. } }

\begin{document}

\maketitle
\thispagestyle{empty} 
 \begin{abstract}
We study a model of opinion dynamics introduced by Krause: each
agent has an opinion represented by a real number, and updates its
opinion by averaging all agent opinions that differ from its own
by less than 1. We give a new proof of convergence into clusters
of agents, with all agents in the same cluster holding the same
opinion. We then introduce a particular notion of equilibrium
stability and provide lower bounds on the inter-cluster distances
at a stable equilibrium. To better understand the behavior of the
system when the number of agents is large, we also introduce and
study a variant involving a continuum of agents, obtaining partial
convergence results and lower bounds on inter-cluster distances,
under some mild assumptions.
\end{abstract}


\section{Introduction}\label{sec:intro}

There has been an increasing interest in recent years in the study
of multi-agent systems \jnt{where agents interact according to
simple local rules, resulting in a possibly coordinated global
behavior.} In a prominent paradigm \JNT{dating back to
\cite{deGroot} and \cite{Tsitsiklis:84phdthesis},} each agent
maintains a value which it updates by taking a linear, and
\jnt{usually} convex combination of other agents' values; see
e.g., \cite{Tsitsiklis:84phdthesis, JadbabaieLinMorse:2003,
Moreau:2005, BlondelHendrickxOlshevskyTsitsiklis:2005,
HendrickxBlondel:2006_MTNS}, and \cite{OlfatiSaberFaxMurray:2006,
RenBeardAtkins:2007} for surveys. The interactions between agents
are generally not all-to-all, but are described by an
interconnection topology. In some applications, this topology is
fixed, but several studies consider the more intriguing case of
changing topologies. For example, in Vicsek's swarming model
\cite{VicsekCzirolBenjacobCohenSchchet:1995}, animals are modeled
as agents that move on the two-dimensional plane. All agents have
the same speed but possibly different headings, and at each
time-step they update their headings by averaging the headings of
those agents that are sufficiently close to them. When the
topology depends on the combination of the agent states, \jnt{as
in Vicsek's model,} an analysis that takes this dependence into
account can be difficult. For this reason, the sequence of
topologies is often treated as exogenous (see e.g.
\JNT{\cite{PDCbook,JadbabaieLinMorse:2003,Moreau:2005}),} with a
few notable exceptions \cite{CuckerSmale:2005, CuckerSmale:2007,
JusthKrishnaprasad:2004}. For instance, the authors of
\cite{CuckerSmale:2005} consider a variation of the model studied
in \cite{JadbabaieLinMorse:2003}, in which communications are
all-to-all, but with the relative importance given by one agent to
another weighted by the distance separating the agents. They
provide conditions under which the agent headings converge to a
common value and the distance between any two agents converges to
a constant. The same authors relax the all-to-all assumption in
\cite{CuckerSmale:2007}, and \jt{study} communications restricted
to arbitrarily changing but connected topologies.

We consider here a simple discrete-time system involving
endogenously changing topologies, and analyze it while taking
explicitly into account the dependence of the topology on the
system state. The \jnt{\modif{discrete-agent}} model is as
follows. There are $n$ agents, and every agent $i$
($i=1,\ldots,n$), maintains a real value $x_i$. These values are
synchronously updated according to
\begin{equation}\label{eq:def_discrete-time_system}
\modif{ x_i(t+1) =
\frac{\sum_{j:|x_i(t)-x_j(t)|<1}x_j(t)}{\sum_{j:|x_i(t)-x_j(t)|<1}1}.}
\end{equation}
Two agents $i$, $j$ for which $|x_i(t)-x_j(t)|<1$ are said to be
\emph{neighbors} or \emph{connected} (at time $t$). Note that with
this definition, an agent is always its own neighbor. Thus, in
this model, each agent updates its value by computing the average
of the values of its neighbors. In the sequel, we usually refer to
the agent values as  ``opinions,'' and sometimes as ``positions.''

\begin{figure}
\centering
\begin{tabular}{cc}
\includegraphics[scale=.29]{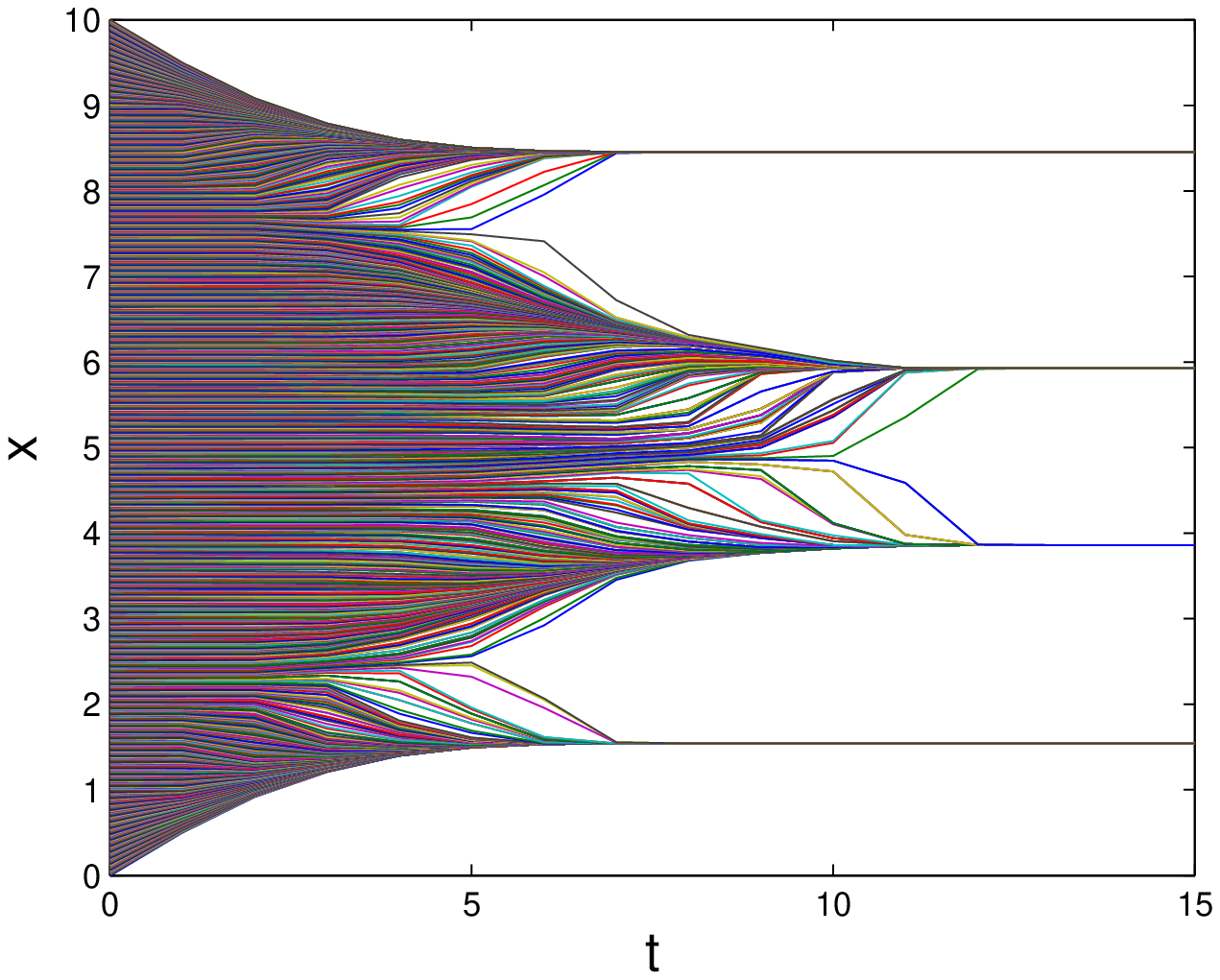}&
\includegraphics[scale=.29]{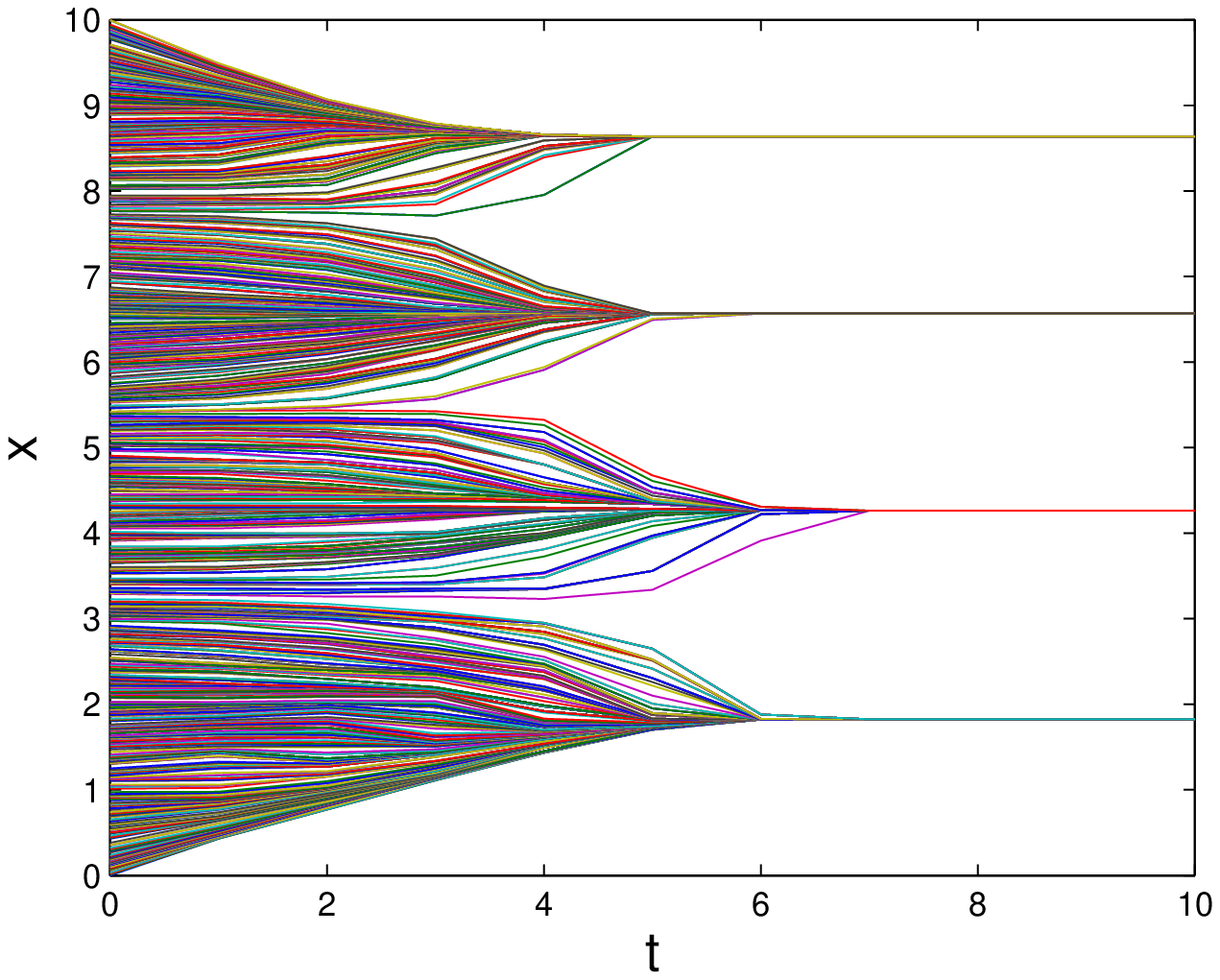}\\(a)&(b)
\end{tabular}

\caption{Time evolution of 1000 agent opinions, according to the
model (\ref{eq:def_discrete-time_system}). Initial opinions are
either uniformly spaced  (case (a)) or chosen at random (case
(b)), on an interval of length $10$. In both cases, opinions
converge to limiting values (``clusters'') that are separated from
each other by much more than the interaction radius, which was set
to 1.}\label{fig:example_2R}
\end{figure}

The model \jt{\eqref{eq:def_discrete-time_system}} was introduced
by Krause \cite{Krause:1997} to capture the dynamics of opinion
formation. Values represent opinions on some subject, and an agent
considers another agent as \quotes{reasonable} if their opinions
differ by less than 1\footnote{\modrev{In Krause's initial
formulation, all opinions belong to $[0,1]$, and an agent
\JNT{considers} another one as reasonable if their opinions differ
by less than a pre-defined parameter $\epsilon$.}}. Each agent
thus updates its opinion by computing the average of the opinions
it finds \quotes{reasonable}. This system is also sometimes
referred to as the Hegselmann-Krause model, following
\cite{HegselmannKrause:2002}. It has been abundantly studied in
the literature \cite{Krause:1997, Krause:2000, Lorenz:2005,
Lorenz:2006}, and displays some peculiar properties that have
remained unexplained. For example, it has been experimentally
observed that opinions initially uniformly distributed on an
interval tend to converge to clusters of opinions separated by a
distance slightly larger than 2, as shown in Figure
\ref{fig:example_2R}. In contrast, presently available results can
only prove convergence to clusters separated by at least 1. An
explanation of the inter-cluster distances observed for this
system, or a proof of a nontrivial lower bound is not available.

Inter-cluster distances larger than the interaction radius (which
in our case was set to 1) have also been observed by Deffuant et
al. \cite{DeffuantNeauAmblardWeisbuch:2000} for a \JNT{related}
\modrev{ stochastic model,} often referred to as the
Deffuant-Weisbuch model. In that model, two randomly selected
agents update their opinions at any given time step. If their
opinions differ by more than a certain threshold, their opinions
remain unchanged; otherwise, each agent moves to a new opinion
which is a weighted average of its previous opinion and that of
the other agent. \JNT{Thus, the Krause and Deffuant-Weisbuch
models rely} on the same idea of bounded confidence, but differ
\JNT{because one is} stochastic while the other is deterministic.
Besides, \JNT{Krause's} model involves simultaneous interactions
between potentially all agents, while the interactions in
\JNT{the} Deffuant-Weisbuch model are pairwise. Despite these
differences, the behavior of these two systems \JNT{is similar,
including inter-cluster distances significantly larger than the
interaction radius.} The behavior of \JNT{the} Deffuant-Wesibuch
model --- and in particular the final positions of the clusters
--- \modrev{has also been studied by considering a continuous
density approximating the discrete distribution of agents, and
examining the partial differential equation describing the
evolution of this density \cite{Ben-naimKrapivskyRedner:2003,
Ben-naimKrapivskyVasquezRedner:2003}.} Other models, involving
either discrete or continuous time, and \jt{finitely} or \modrev{
infinitely many agents, have also been proposed \cite{Urbig:2003
,FortunatoLatoraPluchinoRapisarda:2005, BenNaim:2005}}. For a
survey, see for example \cite{Lorenz:2007}.

The model that we consider also \JNT{has} similarities with
certain rendezvous algorithms (see, e.g.,
\cite{LinMorseAnderson:2003}) in which the objective is to have
all agents meet at a single point. Agents are considered neighbors
if their positions are within a given radius $R$. The update rules
satisfy two conditions. First, when an agent moves, its new
position is a convex combination of its previous \jt{position and
the positions} of its neighbors. Second, if two agents are
neighbors, they remain neighbors after updating their positions.
This ensures that an initially connected set of agents is never
split into smaller groups, so that all agents can indeed converge
to the same point.

In this paper, we start with a simple convergence proof for the
model \eqref{eq:def_discrete-time_system}. We then introduce a
particular notion of equilibrium stability, involving a robustness
requirement when an equilibrium is perturbed by introducing an
additional agent, and prove that an equilibrium is stable if and
only if \jnt{all inter-cluster distances are above a certain
nontrivial lower bound.}  We observe experimentally that the
probability of converging to a stable equilibrium increases with
the number of agents. To better understand the case of a large
numbers of agents, we introduce and study a variation of the
model, which involves a continuum of agents \jnt{(the
``continuous-agent'' model).} We give partial convergence results
and provide a lower bound on the inter-cluster distances at
equilibrium, under some \jnt{regularity} assumptions. We also show
that for a large number of discrete agents, the behavior of the
discrete-agent model  indeed approximates the continuous-agent
model.

Our continuous-agent model, first introduced in
\cite{BlondelHendrickxTsitsiklis:2007ECC}, is obtained by indexing
the agents by a real number instead of an integer. It is
equivalent to the so-called \quotes{discrete-time density based
Hegselmann-Krause model} proposed independently in
\cite{Lorenz:2007}, which is in turn similar to a model presented
in \cite{FortunatoLatoraPluchinoRapisarda:2005} in a
continuous-time setup. Furthermore, our model can also be viewed
as the limit, as the number of discrete opinions tends to
infinity, of the \quotes{interactive Markov chain model}
introduced by Lorenz \cite{Lorenz:2006}; in the latter model,
there is a continuous distribution of agents, but the
\jnt{opinions take values in a discrete set.}

We provide an  analysis of the discrete-agent model
(\ref{eq:def_discrete-time_system}) in Section
\ref{sec:discr_time_discr_agents}. We then consider the
continuous-agent model in Section \ref{sec:discr-time_cont_agent}.
We study the relation between these two models in Section
\ref{sec:discr-time_link}, and we end with concluding remarks and
open questions, in Section \ref{sec:concl}.

\section{The discrete-agent model}\label{sec:discr_time_discr_agents}

\subsection{Basic properties and convergence}

We begin with \modrev{a presentation of} certain basic properties
of the discrete-agent model (\ref{eq:def_discrete-time_system}),
\modrev{most of which have already been proved in
\cite{Krause:2000, Lorenz:2005, HegselmannKrause:2002}}.

\begin{prop}[Lemma 2 in \cite{Krause:2000}]\label{prop:order_preservation}
Let $\prt{x(t)}$ be a sequence of vectors in
$\Re^n$ evolving according to (\ref{eq:def_discrete-time_system}).
The order of opinions is preserved: if $x_i(0) \leq x_j(0)$, then
$x_i(t)\leq x_j(t)$ for all~$t$.
\end{prop}
\begin{IEEEproof}
We use induction. Suppose that $x_i(t)\leq x_j(t)$. Let $N_i(t)$
be the set of agents connected to $i$ and not to $j$, $N_j(t)$ the
set of agents connected to $j$ and not to $i$, and $N_{ij}(t)$ the
set of agents connected to both $i$ and $j$, at time $t$. We
assume here that these sets are \JNT{nonempty,} but our argument can
easily be adapted if some of them are empty. For any $k_1 \in
N_i(t)$, $k_2 \in N_{ij}(t)$, and $k_3 \in N_j(t)$, we have
$x_{k_1}(t)\leq x_{k_2}(t)\leq x_{k_3}(t)$. Therefore, $\bar
x_{N_i}\leq \bar x_{N_{ij}} \leq \bar x_{N_j}$, where $\bar
x_{N_i},\bar x_{N_{ij}},\bar x_{N_j}$, respectively, is the
average of $x_k(t)$ for $k$ in the corresponding set. \modrev{ It
follows from (\ref{eq:def_discrete-time_system}) that
\begin{equation*}
x_i(t+1) = \frac{\abs{N_{ij}}\bar x_{N_{ij}} + \abs{N_{i}}\bar
x_{N_{i}}}{\abs{N_{ij}}+ \abs{N_{i}}} \leq \jnt{\bar x_{N_{ij}}},
\end{equation*}
and
\begin{equation*}
x_j(t+1)=  \frac{\abs{N_{ij}}\bar x_{N_{ij}} + \abs{N_{j}}\bar
x_{N_{j}}}{\abs{N_{ij}}+ \abs{N_{j}}} \geq \bar x_{N_{ij}},
\end{equation*}}
where we use $\abs{A}$ to denote the cardinality of a set $A$.
\end{IEEEproof}
In light of this result, we will assume in the sequel,
without loss of generality,
that the initial opinions are sorted: if $i<j$ then $x_i(t)\leq
x_j(t)$. The next Proposition follows immediately from the
definition of the model.

\begin{prop}\label{prop:monoton_and_separation}
Let $\prt{x(t)}$ be a sequence of
vectors in $\Re^n$ evolving according to
(\ref{eq:def_discrete-time_system}), and such that $x(0)$ is
sorted, i.e., if $i<j$, then $x_i(0) \leq x_j(0)$. The smallest opinion
$x_1$ is nondecreasing with time, and the largest opinion $x_n$ is
nonincreasing with time. Moreover, if at some time the distance
between two consecutive agent opinions $x_i(t)$ and $x_{i+1}(t)$
is larger than or equal to 1 it remains so for all subsequent
times $t'\geq t$, so that the system can then be decomposed into
two independent subsystems containing the agents $1,\dots,i$, and
$i+1,\dots,n$, respectively.
\end{prop}

Note that unlike other related models \modrev{as \JNT{the}
Deffuant-Weisbusch model \cite{DeffuantNeauAmblardWeisbuch:2000}
or the continuous-time model in \cite{Hendrickx:2008phdthesis}},
the average of the opinions is not necessarily preserved, and the
\quotes{variance} (sum of squared differences from the average)
may occasionally increase. \modif{See
\cite{Hendrickx:2008phdthesis} for examples with three and eight
agents respectively.} \modrev{The convergence of
(\ref{eq:def_discrete-time_system}) has already been established
in the literature (see \cite{Dittmer:2001,Lorenz:2005}), and
\JNT{is also easily deduced from the convergence results for the
case of exogenously determined connectivity sequences (see e.g.,
\cite{Moreau:2005, BlondelHendrickxOlshevskyTsitsiklis:2005,
HendrickxBlondel:2006_MTNS, Lorenz:2005}), an approach that
extends to the case of higher-dimensional opinions.} We present
here a simple alternative proof, which \JNT{exploits the
particular dynamics we are dealing with.}}

\begin{thm}\label{thm:conv_discr-time_discr-agent}
If $x(t)$ evolves according to
(\ref{eq:def_discrete-time_system}), then for every $i$, $x_i(t)$
converges to a limit $x_i^*$ in finite time.  Moreover, for any
$i,j$, we have either $x_i^*=x_j^*$ or $\abs{x_i^*-x_j^*}\geq 1$.
\end{thm}
\begin{IEEEproof}
Since $x(0)$ is assumed to be sorted, the opinion $x_1$ is
nondecreasing and bounded above by $x_n(0)$. As a result, it
converges to a value $x_1^*$. Let $p$ be the highest index for
which $x_p$ converges to $x_1^*$.

We claim that if $p<n$, there is a time \jnt{$t$ such that
$x_{p+1}(t)-x_p(t) \geq 1$.}   Suppose, to
obtain a contradiction, that the claim does not hold, i.e., that
$x_{p+1}(t)-x_p(t)$ is \jnt{always} smaller than 1.
Fix some
$\epsilon>0$ and a time after which the distance of $x_i$ from
$x_1^*$, for $i=1,\dots,p$, is less than $\epsilon$. Since $x_{p+1}$
does not converge to $x_1^*$, there is a further time at
which $x_{p+1}$ is larger than $x_1^* + \delta$ for some \JNT{$\delta >
0$.} For such a time $t$, \modrev{$x_p(t+1)$ is at least
\begin{equation*}
\frac{1}{p+1}\prt{\sum_{i=1}^{p+1}x_i(t)}\geq
\frac{1}{p+1}\prt{\jnt{p(x_1^*-\epsilon) + (x_1^*+\delta) }},
\end{equation*}}
which is larger than $x_1^* + \epsilon$ if $\epsilon$ is chosen
sufficiently small. \jnt{This however contradicts the requirement
that $x_p$ remain within $\epsilon$ from $x_1^*$. This
contradiction shows that there exists a time $t$ at which
$x_{p+1}(t)-x_p(t) \geq 1$. Subsequent to that time, using also
Proposition \ref{prop:monoton_and_separation}, $x_p$ cannot
increase and $x_{p+1}$ cannot decrease, so that the inequality
$x_{p+1}-x_p \geq 1$ continues to hold forever. In particular,
agents $1,\ldots,p$ will no more interact with the remaining
agents. Thus, if $p<n$, there will be some finite time after which
the agents $p+1,\dots,n$ behave as an independent system, to which
we can apply the same argument. Continuing recursively, this
establishes the convergence of all opinions to limiting values
that are separated by at least 1.}

It remains to prove that  convergence takes place in finite
time. Consider the set of agents converging to a particular
limiting value. It follows from the argument above that
there is a time after which none of them is connected to any agent
outside  that set. Moreover, since they converge to a common value,
they eventually get sufficiently close so
that they are all connected to each other. When this happens,
they all compute the same average, reach the same opinion at
the next time step, and keep this opinion for all subsequent
times. Thus, they converge in finite time.
Finite time convergence for the entire systems follows because the
number of agents is finite.
\end{IEEEproof}

We will refer to the limiting values to which opinions converge as
\emph{clusters}. With some abuse of terminology, we will also
refer to a set of agents whose opinions converge to a common value
as a cluster.

It can be shown that the convergence time is bounded above by some
constant $c(n)$ that depends only on $n$. On the other hand, an
upper bound that is independent of $n$ is not possible, even if
all agent opinions lie in the interval $[0,L]$ \jt{for} a fixed
$L$. To see this, consider $n$ agents, with $n$ odd, one agent
initially placed at 1, and $(n-1)/2$ agents initially placed at
$0.1$ and $1.9$. All agents will converge to a single cluster at
1, but the convergence time increases to infinity as $n$ grows.

We note that the convergence result in Theorem
\ref{thm:conv_discr-time_discr-agent} does not hold if we consider
the same model but with a countable number of agents. Indeed,
consider a countably infinite number of agents, all with positive
initial opinions. Let $m(y)$ be the number of agents having an
initial opinion $y$. \modrev{Suppose that $\alpha\in(1/2,1)$, and
consider an initial condition for which $m(0) = 0$, $m(\alpha) =
1$, $m(\alpha(k + 1)) = m(\alpha k) + 3m(\alpha(k - 1))$ for every
integer $k>1$, and $m(y)=0$ for every other value of $y$.} Then,
the update rule (\ref{eq:def_discrete-time_system}) implies that
$x_i(t + 1) = x_i(t) + \alpha/2$, for every agent $i$ and time
$t$, and convergence fails to hold. A countable number of agents
also admits equilibria where the limiting values are separated by
less than $1$. An example of such an equilibrium is obtained by
considering one agent at every integer multiple of $1/2$.

\jnt{We also note that equilibria in which clusters are separated
by less than 1 become possible when opinions are elements of a
manifold, instead of the real line. For example, suppose that
opinions belong to $[0,2\pi)$ (identified with elements of the
unit circle), and that two agents are neighbors if and only if
$|x_i-x_j\ (\mbox{mod}\ 2\pi)|<1$. If every agent updates its
angle by moving to the average of its neighbors' angles, it can be
seen that an initial configuration with $n$ agents located at
angles $2\pi k/n$, $k=0,\ldots,n-1$, is an equilibrium. Moreover,
more complex equilibria also exist. Convergence has been
experimentally observed for models of this type, but no proof is
available.}

\subsection{Experimental observations}

Theorem \ref{thm:conv_discr-time_discr-agent} states that opinions
converge to clusters separated by at least 1. Since the smallest
and largest opinions are nondecreasing and nonincreasing,
respectively, it follows that opinions initially confined to an
interval of length $L$ can converge to at most $\jnt{\ceil{L}} +1$
clusters. It has however been observed in the literature that the
distances between clusters are usually significantly larger than 1
(see \cite{Krause:2000, Lorenz:2006}, and Figure
\ref{fig:example_2R}), resulting in a number of clusters that is
significantly smaller than the upper bound of $\ceil{L}+1$.
\modrev{To further study this phenomenon, we analyze below
different experimental results, similar to those in
\cite{Lorenz:2006}.

Figure \ref{fig:evol number cluster} shows the dependence on $L$
of the cluster number and positions, for the case of a large
number of agents and initial opinions that are uniformly spaced
\jnt{on an interval of length $L$.}} Such incremental analyses
also appear in the literature for various similar systems
\cite{Ben-naimKrapivskyRedner:2003, Hegselmann:2004, Lorenz:2006,
Lorenz:2007}. We see that the cluster positions tend to change
with $L$ in a piecewise continuous (and sometimes linear) manner.
The discontinuities correspond to the emergence of new clusters,
or to the splitting of a cluster into two smaller ones. The number
of clusters tends to increase linearly with $L$, with a
coefficient slightly smaller than $1/2$, corresponding to an
inter-cluster distance slightly larger than $2$. Note however that
this evolution is more complex than it may appear: Irregularities
in the distance between clusters and in their weights can be
observed for growing $L$, as already noted in \cite{Lorenz:2006}.
Besides, for larger scale simulations ($L=1000,n=10^6$), a small
proportion of clusters take much larger or much smaller weights
than the others, and some inter-cluster distances are as large as
4 or as small as 1.5. These irregularities could be inherent to
the model, but may also be the result of the particular
discretization chosen or of the accumulation of numerical errors
in a discontinuous system.

Because no nontrivial lower bound is available to explain the
observed inter-cluster distances in Krause's model, we start with
three observations that can lead to some partial understanding. In
fact, the last observation will lead us to a formal stability
analysis, to be developed in the next subsection.

\begin{figure}
\centering
\includegraphics[scale=.25]{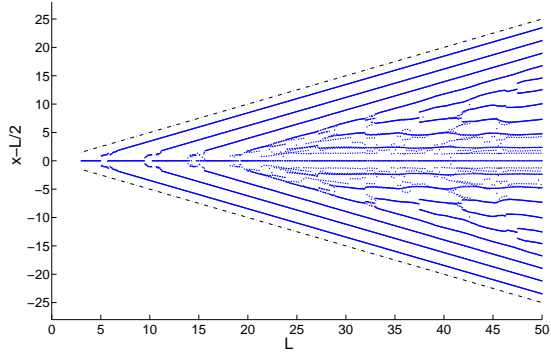}
\caption{Locations of the different clusters at equilibrium, as a
function of $L$, for $5000 L$ agents whose initial opinions are
uniformly spaced on $\jt{[0,L]}$, \jt{represented in terms of
their distance from $L/2$.} The dashed lines \jt{correspond to}
the endpoints \jt{$0$ and $L$} of the initial opinion
distribution. Similar results are obtained if the initial opinions
are chosen at random, with a uniform distribution.}\label{fig:evol
number cluster}
\end{figure}

\emph{(a)} We observe from Figure \ref{fig:evol number cluster}
that the minimal value of $L$ that leads to multiple clusters is
approximately $5.1$, while Theorem
\ref{thm:conv_discr-time_discr-agent} only requires that this
value be at least 1. This motivates us to address the question of
whether a more accurate bound can be derived analytically. Suppose
that there is an odd number of agents whose initial opinions are
uniformly spaced on $[0,L]$. An explicit calculation shows that
all opinions belong to an interval
$[\frac{1}{2}-O(\frac{1}{n}),L-\frac{1}{2}+O(\frac{1}{n})]$ after
one iteration, and to an interval $[\frac{11}{12}-O(\frac{1}{n}),L
- \frac{11}{12}+O(\frac{1}{n})]$ after two iterations.
Furthermore, by Proposition \ref{prop:monoton_and_separation}, all
opinions must subsequently remain inside these intervals. On the
other hand, note that with an odd number of agents, there is one
agent that always stays at $L/2$. Thus, if all opinions eventually
enter the interval $(L/2-1,L/2+1)$, then there can only be a
single cluster. This implies that there will be a single cluster
if $\jt{L- \frac{11}{12} +O(\frac{1}{n}) < L/2 +   1} $ , that is,
if $L < \frac{23}{6} -O(\frac{1}{n}) \simeq 3.833$. This bound is
smaller than the \jt{experimentally} observed value of about 5.1.
It can be further improved by carrying out explicit calculations
of the smallest position after a further number of iterations.
Also, as long as the number of agents is sufficiently large, a
similar analysis is possible if the number of agents is even, or
in the presence of random initial opinions.

\emph{(b)} When $L$ is sufficiently large, Figure \ref{fig:evol
number cluster} shows that the position of the leftmost clusters
becomes independent of $L$. This can be explained by analyzing the
propagation of information: at each iteration, an agent is only
influenced by those opinions within distance $1$ of its own, and
its opinion is modified by less than $1$. So, information is
propagated by at most a distance 2 at every iteration. For the
case of  uniformly spaced initial opinions on $[0,L]$, with $L$
large, the agents with initial opinions close to 0 behave, at
least in the first iterations, as if opinions were initially
distributed uniformly on $[0,+\infty)$. Moreover, once a group of
opinions is separated from other opinions by more than $1$, this
group becomes decoupled. Therefore, if the agents with initial
opinions close to 0 become separated from the remaining agents in
finite time, their evolution under a uniform initial distribution
on $[0,L]$ for a sufficiently large $L$ is the same as in the case
of a uniform initial distribution on $[0,+\infty)$.

\begin{figure}
\centering
\includegraphics[scale=.5]{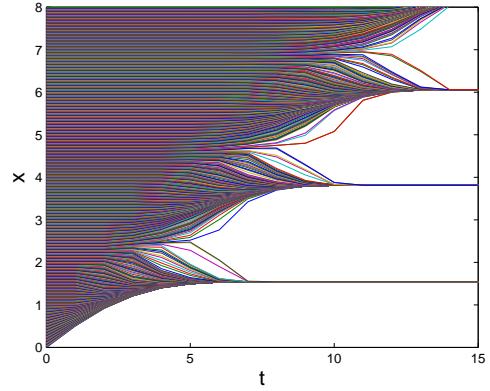}
\caption{Time evolution when the initial opinions are uniformly
spaced on a semi-infinite interval, with a density of 100 per unit
length. Groups of agents become separated from the remaining
agents, and converge to clusters separated by
approximately~2.2.}\label{fig:semi-inf}
\end{figure}

We performed simulations with initial opinions uniformly spaced on
$[0,\infty)$, \modrev{as in \cite{Lorenz:2006}}. We found  that
every agent eventually becomes connected with a finite number of
agents and disconnected from the remaining agents. The groups
formed then behave independently and converge to clusters. As
shown in Figure \ref{fig:semi-inf}, the distances between two
consecutive clusters \modrev{are close to $2.2$. These distances
partially explain the evolution of the number of clusters (as a
function of $L$) shown in Figure \ref{fig:evol number cluster}.
However, a proof of these observed properties is not available,
and it is unclear \JNT{whether} the successive inter-cluster
distances \JNT{possess some regularity or convergence
properties.}}

\begin{figure}
\centering
\includegraphics[scale=.5]{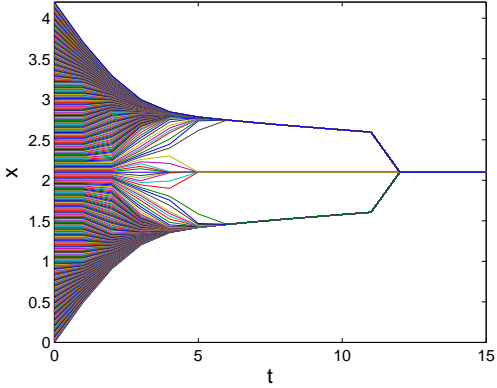}
\caption{Example of a temporary \quotes{meta-stable} state.
Initially, two groups are formed that do not interact with each
other, but they both interact with a small number of agents lying
in between. As a result, the distance separating the two groups
decreases slowly and eventually becomes smaller than 1. At that
point, the groups attract each other directly and merge into a
single cluster.} \label{fig:metastable_discr_time}
\end{figure}

\emph{(c)} A last observation that leads to a better understanding
of the size of the inter-cluster distances is the following.
Suppose that $L$ is just below the value at which two clusters are
formed, and note the special nature of the resulting evolution,
shown in Figure \ref{fig:metastable_discr_time}. The system first
converges to a \quotes{meta-stable state} in which there are two
groups, separated by a distance slightly larger than 1, and which
therefore do not interact directly with each other. The two groups
are however slowly attracted by some isolated agents located in
between; furthermore, these isolated agents are being pulled by
both of these groups and remain at the weighted average of the
opinions in the two groups. Eventually, the distance between the
two groups becomes smaller than 1, the two groups start attracting
each other directly, and merge into a single cluster.
(\modrev{This corresponds to one of the slow convergence phenomena
observed in \cite{Lorenz:2006}.}) The initial convergence towards
a two-cluster equilibrium is thus made impossible by the presence
of a few agents in between. Moreover, the number of these isolated
agents required to destabilize a meta-stable state can be
arbitrarily small compared to the number of agents in the two
groups. On the other hand, this phenomenon will not arise  if the
two clusters are separated by a sufficiently large distance. For
example, if the distance between the two groups is more than 2, no
agent can be simultaneously connected to both groups. This
suggests that, depending on the distance between clusters, some
equilibria are stable with respect to the presence of a small
number of additional agents, while \jt{some} are not.

\subsection{Stability with respect to a perturbing agent}\label{sec:stab_discr_agents}

In this section, we introduce a notion of equilibrium stability,
motivated by the last observation in the preceding subsection. We
first generalize the model (\ref{eq:def_discrete-time_system}), so
that each agent $i$ has an associated weight $w_i$ and updates its
opinion according to the weighted discrete-agent model
\begin{equation}\label{eq:weighted_1D_model}
x_i(t+1) = \frac                     %
{\sum_{j:\abs{x_i(t)-x_j(t)}<1}w_j x_j(t)}
{\sum_{j:\abs{x_i(t)-x_j(t)}<1}w_j }.
\end{equation}
It can be verified that the convergence results in Theorem
\ref{thm:conv_discr-time_discr-agent} and the properties proved in
Propositions \ref{prop:order_preservation} and
\ref{prop:monoton_and_separation} continue to hold. We will use
the term \emph{weight of a cluster} to refer to the sum of the
weights of all agents in the cluster. Observe that if a number $w$
of agents in system (\ref{eq:def_discrete-time_system}) have the
same position, they behave as a single agent with weight $w$ in
the model \eqref{eq:weighted_1D_model}. \jnt{This correspondence
can also be reversed, so that \eqref{eq:weighted_1D_model} can be
viewed as a special case of (\ref{eq:def_discrete-time_system}),
whenever the weights $w_i$ are integer,} \jt{or more generally,
rational numbers.}

Let $\bar x$ be a vector of agent opinions at equilibrium. Suppose
that we add a perturbing agent indexed by 0, with weight $\delta$
and initial opinion $\tilde x_0$,  that we let the system evolve
again, until it converges to a new, perturbed equilibrium, and
then remove the perturbing agent. The opinion vector $\bar x'$ so
obtained is again an equilibrium. We define $\Delta_{\tilde
x_0,\delta} = \sum_{i}w_i\abs{\bar x_i-\bar x_i'}$, which is a
measure of the distance between the original and perturbed
equilibria. We say that $\bar x$ is \emph{stable} if $\sup_{\tilde
x_0}\Delta_{\tilde x_0,\delta}$, the \modif{supremum} of distances
between initial and perturbed equilibria caused by a perturbing
agent of given weight $\delta$, \jnt{converges to zero as}
\modif{$\delta$ vanishes.} Equivalently, an equilibrium is
unstable if a substantial change in the equilibrium can be induced
by a perturbing agent of arbitrarily small weight.

\begin{thm}\label{thm:stab_disc_time_disc_agent}
\JNT{An equilibrium is stable if and only if for any two clusters
$A$ and $B$ with weights $W_A$ and $W_B$, respectively, \JNT{the
following holds: either $W_A=W_B$ and the inter-cluster distance
is greater than or equal to 2; or $W_A\neq W_B$ and the
inter-cluster distance is \emph{strictly} greater than
$1+\frac{\min\prt{W_A,W_B}}{\max\prt{W_A,W_B}}$.} (Note that the
two cases are consistent, except that the second involves a strict
inequality.)}
\end{thm}

\begin{IEEEproof}
We start with an interpretation of the \JNT{strict inequality} in
the statement of the theorem. Consider two clusters $A$ and $B$,
at positions $x_A$ and $x_B$, and let $m=(W_A x_A+W_B
x_B)/(W_A+W_B)$, \jt{which is} their center of mass. Then, an easy
calculation shows that
\begin{equation}\label{eq:condition}
\begin{array}{c}
|x_A-x_B|>1+\frac{\min\prt{W_A,W_B}}{\max\prt{W_A,W_B}}\\

\mbox{if and only if}\\ \max\{|m-x_A|,\, |m-x_B|\}>1\end{array}
\end{equation}

Suppose that an equilibrium $\bar x_0$ satisfies the
\JNT{conditions} in the theorem.  We will show that $\bar x_0$ is
stable. Let us insert a perturbing agent of weight $\delta$. Note
that since $\bar x_0$ is an equilibrium, and therefore the
clusters are at least 1 apart, the perturbing agent is connected
to at most two clusters. If this agent is disconnected from all
clusters, it has no influence, and $\Delta_{\tilde x_0,\delta}
=0$. If it is connected to exactly one cluster $A$, with position
$x_A$ and weight $W_A$, the system reaches a new equilibrium after
one time step, where both the perturbing agent and the cluster
have an opinion $(\tilde x_0 \delta + x_A W_A) / (\delta + W_A)$.
Then,
$$\Delta_{\tilde x_0,\delta}=
\abs{\tilde x_0-x_A}\cdot\frac{\delta}{\delta+W_A} \leq
\frac{\delta}{\delta+W_A},$$ which converges to 0 as $\delta\to
0$. Suppose finally that the perturbing agent is connected to two
clusters $A,B$. This implies that the distance between these two
clusters is less than 2, and since $\tilde x_0$ satisfies the
\JNT{conditions} in the theorem, it must be greater than
$1+\frac{\min\prt{W_A,W_B}}{\max\prt{W_A,W_B}}$. Therefore, using
\eqref{eq:condition}, the distance of  one these clusters from
their center of mass $m$ is greater than 1. The opinion of the
perturbed agent after one iteration is within $O(\delta)$ from
$m$, while the two clusters only move by an $O(\delta)$ amount.
Since the original distance between one of the two clusters and
$m$ is greater than 1, it follows that after one iteration, and
when $\delta$ is sufficiently small, the distance of the
perturbing agent from one of the clusters is greater than 1, which
brings us back to the case considered earlier, and again implies
that $\Delta_{\tilde x_0,\delta}$ converges to zero as $\delta$
decreases.

\JNT{To prove the converse, we now suppose} that the distance
between two clusters $A$ and $B$, at positions $x_A$ and $x_B$, is
less than 2, and also less than
$1+\frac{\min\prt{W_A,W_B}}{\max\prt{W_A,W_B}}$. Assuming without
loss of generality that $x_A < x_B$, their center of mass $m$ is
in the interval $(x_B-1,x_A+1)$. Let us fix an $\epsilon>0$ such
that $(m-\epsilon,m+\epsilon)\subseteq (x_B-1,x_A+1)$. Suppose
that at some time $t$ after the introduction of the perturbing
agent we have
\begin{equation}\label{eq:recur_condition_stab_discrete}
\tilde x_0(t) \in \prt{m(t)-\epsilon,m(t)+\epsilon}\subseteq
\prt{x_B(t)-1,x_A(t)+1},
\end{equation}
with $x_B(t) - x_A(t) \geq 1$, where $\tilde x_0(t)$, $x_A(t)$, $
x_B(t)$, and $m(t)$ represent the positions at time $t$ of the
perturbing agent, of the clusters A and B, and of their center of
mass, respectively. One can easily verify that $x_A(t+1) = x_A(t)
+ \abs{\Theta(\delta)} > x_A(t)$, and $x_B(t+1) = x_B(t) -
\abs{\Theta(\delta)}$, so that $x_B(t+1) -x_A(t+1) < x_B(t)
-x_A(t)$, and $\prt{m(t+1)-\epsilon,m(t+1)+\epsilon}\subseteq
\prt{x_B(t+1)-1,x_A(t+1)+1}.$

Moreover, observe that if $\delta$ were 0, we would have $\tilde
x_0 (t+1) = m(t)$. For $\delta \not =0$, $\tilde x_0 (t+1)$ is
close to $m(t)$, and we have $\tilde x_0(t+1) = m(t) + O(\delta)$.
Since  $$m(t+1) = \frac{W_A x_A(t+1) + W_B x_B(t+1)}{W_A+W_B} =
m(t) + O(\delta),$$ we obtain $\abs{m(t+1) - m(t)} = O(\delta)$,
and therefore $\tilde x_0(t+1) \in
(m(t+1)-\epsilon,m(t+1)+\epsilon)$, as long as $\delta$ is
sufficiently small with respect to $\epsilon$.

We have shown that if $\tilde x_0(0)=\tilde x_0$ is chosen so that
the condition (\ref{eq:recur_condition_stab_discrete}) is
satisfied for $t=0$, and if $\delta$ is sufficiently small, the
condition (\ref{eq:recur_condition_stab_discrete}) remains
satisfied as long as $x_B(t)-x_A(t) \geq 1$. The perturbing agent
remains thus close to the center of mass, attracting both
clusters, until at some time $t^*$ we have $x_B(t^*)-x_A(t^*) <
1$. The two clusters then merge at the next time step. The result
of this process is independent of the weight $\delta$ of the
perturbing agent, which proves that $\bar x$ is not stable.
Finally, a similar but slightly more complicated argument shows
that $\bar x$ is not stable \modrev{when $|x_A-x_B|= 1 +
\frac{\min\prt{W_A,W_B}}{\max\prt{W_A,W_B}}$, and $|x_A-x_B| <2$.}
\end{IEEEproof}

\begin{figure}
\centering
\includegraphics[scale=.5]{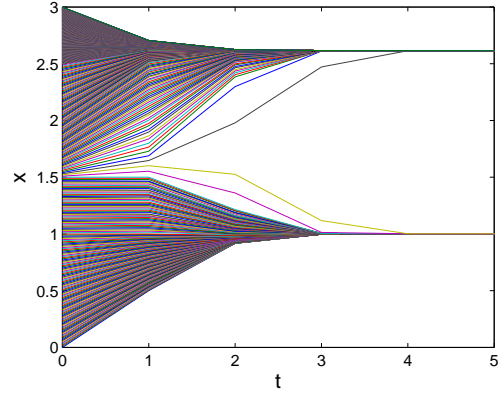}
\caption{Example of convergence to a stable equilibrium where the
clusters are separated by less than 2. The initial distribution of
opinions is obtained by taking 251 uniformly spaced opinions on
$[0,2.5]$ and 500 uniformly opinions on $[2.5,3]$. Opinions
converge to two clusters with 153 and 598 agents, respectively,
that are separated by a distance $1.6138> 1.2559 = 1 +
\frac{153}{598}$. Similar results are obtained when larger number
of agents are used, provided that the initial opinions are
distributed in the same way, i.e, with a density on $[2.5,3]$
which is \jnt{ten} times larger than the density on $[0,2.5]$.
}\label{fig:stab_discr_lessthan2}
\end{figure}

Theorem \ref{thm:stab_disc_time_disc_agent} characterizes the
stable equilibria in terms of a lower bound on the inter-cluster
distances. It allows for inter-cluster distances  at a stable
equilibrium that are smaller than 2, provided that the clusters
have different weights. This is consistent with experimental
observations for certain initial opinion distributions, as shown
in Figure \ref{fig:stab_discr_lessthan2}. On the other hand, for
the frequently observed case of clusters with equal weights,
stability requires the inter-cluster \jt{distances} to be at least 2.
\jnt{Thus, this result comes close to a full explanation of the
observed inter-cluster distances of about 2.2.}

In general, there is no guarantee that the system
(\ref{eq:def_discrete-time_system}) will converge to a stable
equilibrium. (A trivial example is obtained by initializing the
system at an unstable equilibrium, \modrev{such as $x_i(0) =
-\frac{1}{2}$ for half of the agents and $x_i(0) = \frac{1}{2}$
for the other half}). On the other hand, we have observed that for
a given smooth distribution of initial opinions, and as the number
of agents increases, we almost always obtain convergence to a
stable equilibrium. This leads us to  the following conjecture.

\begin{conj}\label{conj:stab_discr_ag_discr_time}
Suppose that the initial opinions are chosen randomly and
independently according to a particular continuous and bounded
probability density function (PDF) with connected support. Then,
the probability of convergence to a stable equilibrium tends to 1,
as the number of agents increases to infinity.
\end{conj}

\begin{figure}
\centering
\begin{tabular}{cc}
\includegraphics[scale=.28]{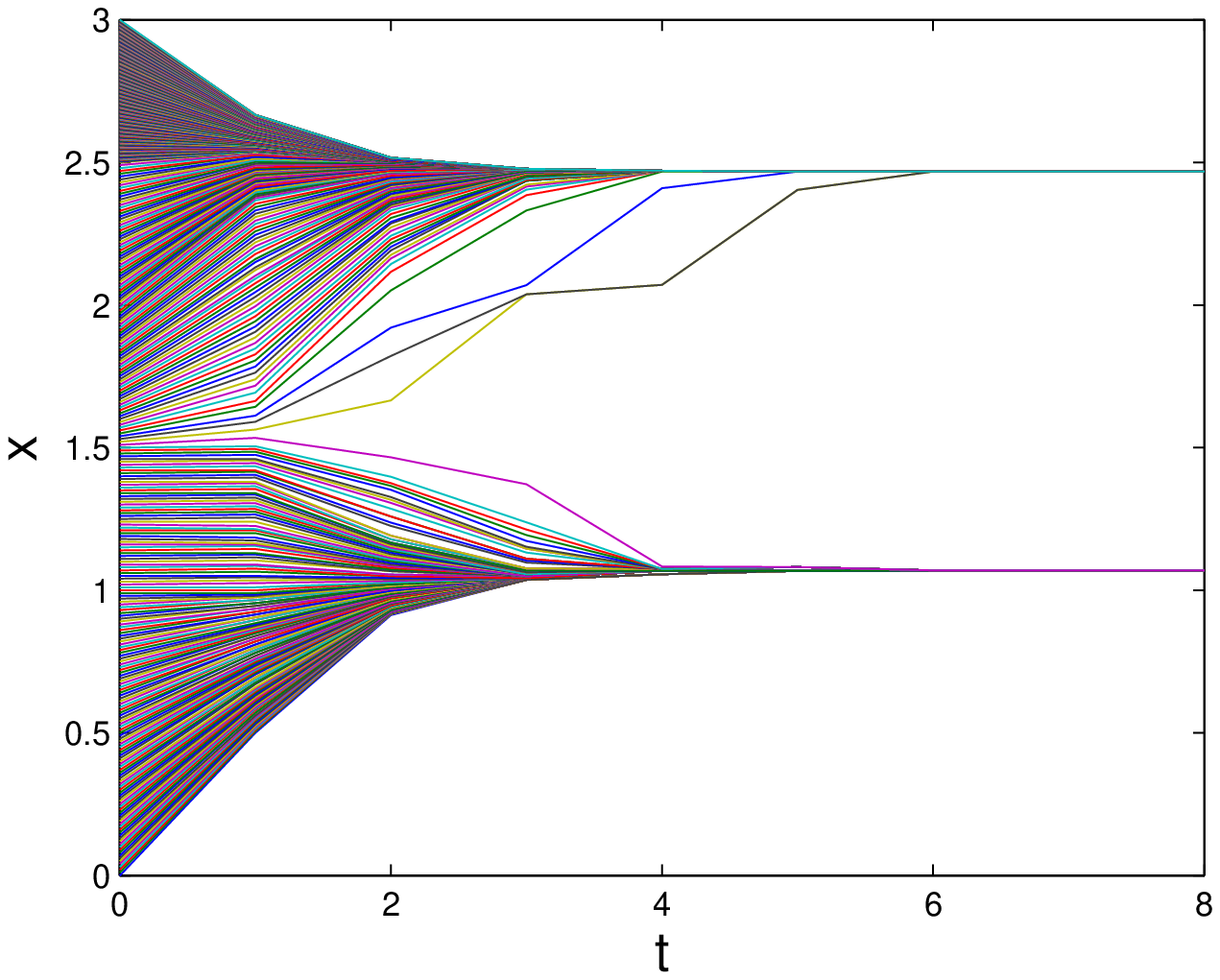}&
\includegraphics[scale=.28]{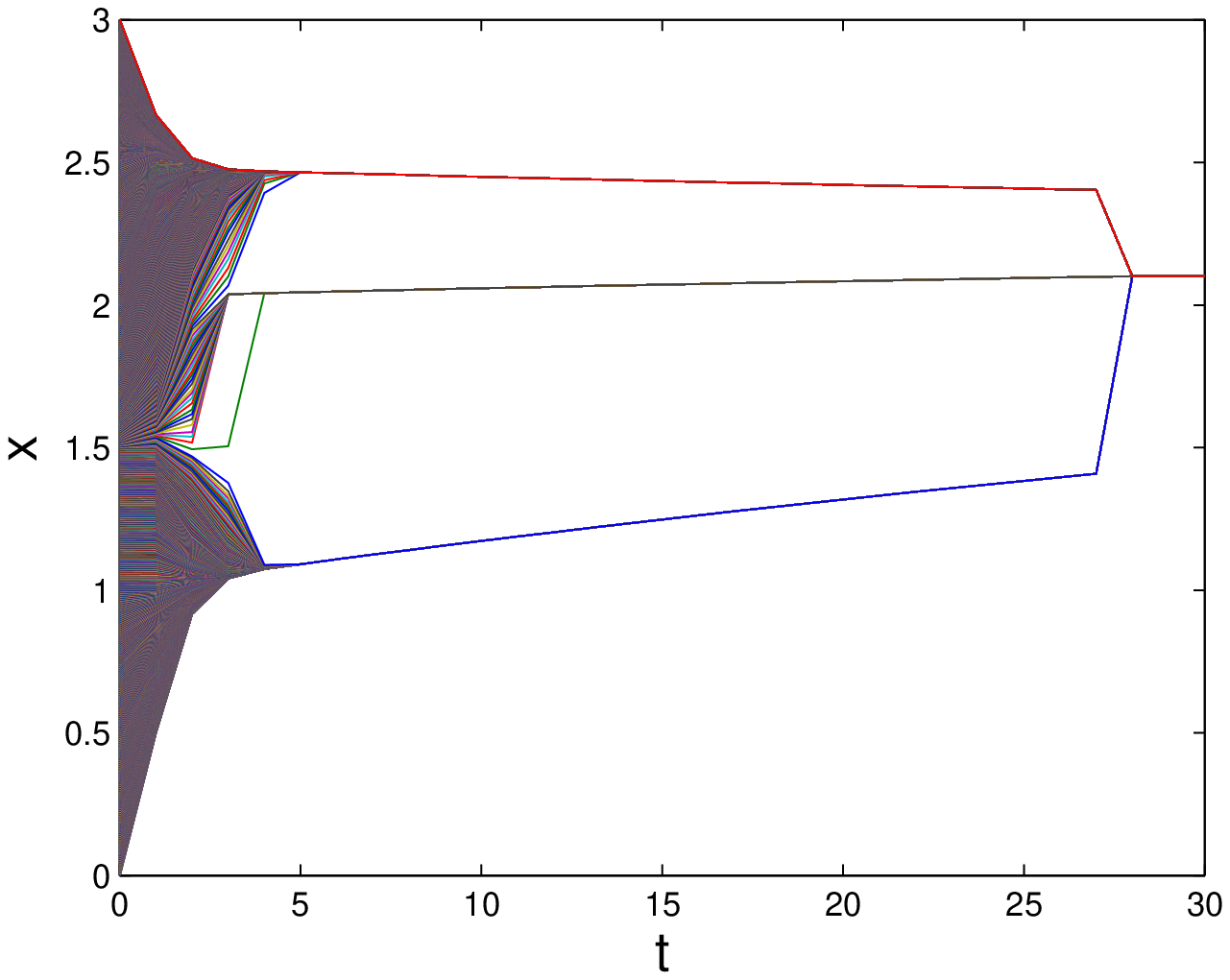}\\(a)&(b)
\end{tabular}

\caption{Time evolution of agent opinions, when initial opinions
are drawn from a common PDF which is larger on the interval
(2.5,3) than on the  interval (0,2.5). In (a), we have 501 agents
and they converge to an unstable equilibrium: the clusters have
respective weights 152 and 349, and their distance is
$1.399<1+\frac{152}{349} \simeq 1.436$. In (b), we have 5001
agents and they converge to a stable equilibrium: we see two
clusters being formed originally, but they are eventually drawn
together by a small number of agents in
between.\label{fig:unstab2stab_n_increases_disc_time}}
\end{figure}

Besides the extensive numerical evidence (see e.g., Figure
\ref{fig:unstab2stab_n_increases_disc_time}), this conjecture is
supported by the intuitive idea that \jt{if} the number of agents
is sufficiently large, whenever two groups of agents start forming
two clusters, there will still be a small number agents in
between, whose presence will preclude convergence to an unstable
equilibrium. The conjecture is also supported by Theorem
\ref{thm:no_conv_2_unstab_cont} in Section
\ref{sec:discr-time_cont_agent}, which deals with a continuum of
agents, together with the results in Section
\ref{sec:discr-time_link} that provide a link between the
discrete-agent and continuous-agent models.

\section{The continuous-agent model}\label{sec:discr-time_cont_agent}

\jnt{The discussion in the previous section indicates that much
insight can be gained by focusing on the case of a large number of
agents. This motivates us to consider a model involving} a
continuum of agents. We use the interval $I=[0,1]$ to index the
agents, and we consider opinions that are nonnegative and bounded
above by a positive constant $L$. We denote by $x_t(\alpha)$ the
opinion of agent $\alpha\in I$ at time $t$. We use $X$ to denote
the set of measurable functions $x:I\to\Re$, and $X_L\subset X$
the set of measurable functions $x:I\to[0,L]$. The evolution of
the opinions is described by
\begin{equation}\label{eq:def_cont_agent_discr_time}
x_{t+1}(\alpha) = \frac{\int_{\beta:(\alpha,\beta)\in
C_{x_t}}x_t(\beta)\, d\beta} {\int_{\beta:(\alpha,\beta)\in C_{x_t}}
d\beta},
\end{equation}
where $C_x\subseteq I^2$ is defined for any $x\in X$ by
\begin{equation*} C_x := \{(\alpha,\beta)\in
I^2:\abs{x(\alpha)-x(\beta)}< 1\}.
\end{equation*}
If the denominator in (\ref{eq:def_cont_agent_discr_time}) is
zero, we use the convention $x_{t+1}(\alpha) = x_t(\alpha)$.
However, since the set of agents $\alpha$ for which this
convention applies has zero measure, we can ignore such agents in
the sequel. \jnt{We assume that $x_0\in X_L$.} We then see that
for every $t> 0$, we have $x_t\in X_L$, so that the dynamics are
well-defined. In the sequel, we denote by $\chi_x$ the indicator
function of $C_x$, that is, $\chi_x(\alpha,\beta) =1$ if
$(\alpha,\beta)\in C_x$, and $\chi_x(\alpha,\beta) =0$ otherwise.

We note that for the same reasons as in the discrete-agent model,
if for some $\alpha$ and $\beta$ we have the relation
$x_t(\alpha)\leq x_t(\beta)$ or $x_t(\alpha) = x_t(\beta)$ at some
$t$, then the same relation continues to hold at all subsequent
times. Furthermore, if $x_0$ only takes a finite number of values,
the continuous-agent model coincides with the weighted
discrete-agent model \eqref{eq:weighted_1D_model}, with the same
range of initial opinions, and where each discrete agent's weight
is set equal to the measure of the set of indices $\alpha$ for
which $x_0(\alpha)$ takes the corresponding value.

\jnt{In the remainder of this section, we will study the
convergence properties of the continuous-agent model, and the
inter-cluster distances at suitably defined stable equilibria.}

\subsection{{Operator formalism}}

To analyze the continuous-agent model
(\ref{eq:def_cont_agent_discr_time}), it is convenient to
introduce a few concepts, extending well known matrix and graph
theoretic tools to the continuous case. By analogy with
interaction graphs in discrete multi-agent systems, we define for
$x\in X$ the {\it adjacency operator} $A_x$, which maps the set
$X$ of measurable functions on $I$ into itself, by letting
\begin{equation*} \prt{A_{x}y}(\alpha) =
\int\chi_x(\alpha,\beta) y(\beta)\, d\beta.
\end{equation*}
Applying this operator can be viewed as multiplying \JNT{$y$} by the
\quotes{continuous adjacency \JNT{matrix}} $\chi_x$, and  using an
extension of the matrix product to the continuous case. We also
define the \modrev{\emph{degree function}}  $d_x:I\rightarrow
\Re^+$, representing the measure of the set of agents to which a
particular agent is connected, by
\begin{equation*}
d_x(\alpha) = \int\chi_x(\alpha,\beta)d\beta = (A_x \1)(\alpha),
\end{equation*}
where $\1:I\to \jnt{\{1\}}$ is the constant function that takes the
value 1 for every $\alpha\in I$. Multiplying a function by the
degree function can be viewed as applying an operator $D_x:X\to
X$ defined by
\begin{equation*}
\prt{D_xy}(\alpha)= d_x(\alpha)y(\alpha) =
\int\chi_x(\alpha,\beta)y(\alpha)\,d\beta.
\end{equation*}
When $d_x$ is positive everywhere, we can also define the operator
$D_x^{-1}$, \jt{which} multiplies a function by $1/d_x$. Finally, we
define the \emph{Laplacian operator} $L_x = D_x - A_x$. It follows
directly from these definitions that $L_x \1 =0$, similar to what
is known for the Laplacian matrix. In the sequel, we also use the
scalar product $\scalprod{x}{y} = \int x(\alpha)y(\alpha)\,d\alpha$.
We now introduce two lemmas to ease the \JNT{manipulation} of these
operators.

\begin{lem}\label{lem:cons:symmetry_of_operators}
The operators defined above are symmetric with respect to the
scalar product: for any $x,y,z\in X$, we have $\scalprod{z}{A_xy}
= \scalprod{A_xz}{y}$, $\scalprod{z}{D_xy} = \scalprod{D_xz}{y}$,
and $\scalprod{z}{L_xy} = \scalprod{L_xz}{y}$.
\end{lem}
\begin{IEEEproof}
The result is trivial for $D_x$. For $A_x$, we have
\begin{equation*}
\begin{array}{lll}
\scalprod{z}{A_xy}&=&\int z(\alpha)\prt{
\int\chi_x(\alpha,\beta)y(\beta)\,d\beta}d\alpha\\
 &=& \int  y(\beta) \prt{
\int \chi_x(\alpha,\beta)z(\alpha)\,d\alpha}d\beta.
\end{array}
\end{equation*}
\modrev{Since $\chi_x(\alpha,\beta) = \chi_x(\beta,\alpha)$ for
all $\alpha,\beta$, this implies
$\scalprod{z}{A_xy}=\scalprod{A_xz}{y}$.} By linearity, the result
also holds for $L_x$ and any other linear combination of those
operators.
\end{IEEEproof}

\begin{lem}\label{lem:cons:scalprod_to_intsquare}
For any $x,y\in X$, we have
\begin{equation*} \scalprod{y}{(D_x \pm A_x)y}  =
\frac{1}{2}\int\chi_x(\alpha,\beta) \prt{y(\alpha) \pm
y(\beta)}^2 d\alpha\, d\beta.
\end{equation*}
In particular, $L_x=D_x-A_x$ is positive semi-definite.
\end{lem}
\begin{IEEEproof}
From the definition of the operators, we have
\begin{equation*}
\scalprod{y}{(D_x\pm A_x) y} = \int\chi_x(\alpha,\beta)
y(\alpha)\prt{y(\alpha)\pm y(\beta)}d\alpha\, d\beta.
\end{equation*}
The right-hand side of this equality can be rewritten as
\begin{equation*}\begin{array}{ll}
&\frac{1}{2}\prt{\int\chi_x(\alpha,\beta)
y(\alpha)\prt{y(\alpha)\pm y(\beta)}d\alpha\, d\beta} \\+&
\frac{1}{2}\prt{\int\chi_x(\beta,\alpha) y(\beta)\prt{y(\beta)\pm
y(\alpha)}d\alpha\, d\beta}.\end{array}
\end{equation*}
\modrev{The symmetry of $\chi_x$ then implies that
$\scalprod{y}{(D_x\pm A_x) y}$ equals
\begin{equation*}
\frac{1}{2}\int\chi_x(\alpha,\beta)\prt{ y(\alpha)^2 \pm
2y(\alpha)y(\beta) + y(\beta)^2} d\alpha\, d\beta,
\end{equation*}
from which the results follows directly.}
\end{IEEEproof}

The update equation (\ref{eq:def_cont_agent_discr_time}) can
be rewritten, more compactly, in the form
\begin{equation}\label{eq:deltax=-dLx} \Delta x_t :=
x_{t+1}-x_t = - D_x^{-1} L_{x_t} x_t, \phantom{a} \text{or}
\phantom{a} D_{x_t}\Delta x_t = - L_{x_t}x_t,
\end{equation}
where the second notation is formally more general as it also
holds on the possibly \JNT{nonempty} zero-measure set on which
$d_x=0$. We say that $x_t\in X_L$ is a \emph{fixed point} of the
system if $\Delta x_t=0$ holds \jnt{almost everywhere (a.e., for
short), that is, except possibly on a zero-measure set.} It
follows from (\ref{eq:deltax=-dLx}) that the set of fixed points
is characterized by the equality $L_x x =0$, a.e. One can easily
see that the set of fixed points contains the set $F:=\{x\in X_L:
x(\alpha)\not = x(\beta) \Rightarrow \abs{x(\alpha)-x(\beta)}\geq
1\}$ of opinion functions taking a discrete number of values that
are at least one apart. Let $\bar F$ be the set of functions
$\jt{x\in X_L}$ for which there exists $s\in F$ such that $s=x$,
a.e. We prove later that $\bar F$ is exactly the set of solutions
to $L_x x =0$, \jnt{a.e.}, and thus the set of fixed points
of~(\ref{eq:deltax=-dLx}).

\subsection{Convergence}

In this section we present some partial convergence results. In
particular, we show that the change $\Delta x_t$ of the opinion
function decays to $0$, and that $x_t$ tends to the set of fixed
points. We begin by proving the decay of a quantity related to
$\Delta x_t$.

\begin{thm}\label{thm:lyapunov_discr_time}
For any initial condition of the system (\ref{eq:deltax=-dLx}),
we have
\begin{equation*}
\sum_{t=0}^\infty \int\chi_{x_t}(\alpha,\beta)\prt{\Delta
x_t(\alpha) +\Delta x_t(\beta)}^2 d\alpha\,  d\beta < \infty.
\end{equation*}
\end{thm}
\begin{IEEEproof}
We consider the nonnegative potential function $V:X\to
\Re^+$ defined by
\begin{equation}\label{eq:def_lyapunov_xLx}
V(x) =\frac{1}{2} \int
\min\prt{1,\prt{x(\alpha)-x(\beta)}^2}d\alpha\, d\beta\geq 0,
\end{equation}
and show that
\begin{equation*}
V(x_{t+1})-V(x_{t})\leq - \scalprod{\Delta x_t}{(A_{x_t} +
D_{x_t})\Delta x_t},
\end{equation*}
which by Lemma \ref{lem:cons:scalprod_to_intsquare} implies the
desired result.

We observe that for every $x,y\in X$, since
$\min\prt{1,\prt{y(\alpha)-y(\beta)}^2}$ is smaller than or equal
to both $1$ and $\prt{y(\alpha)-y(\beta)}^2$, there holds
\begin{equation}\label{eq:cons:upboundVY_cont}
\begin{array}{lll}
V(y) &\leq& \frac{1}{2} \int_{ C_x}
\prt{y(\alpha)-y(\beta)}^2d\alpha\, d\beta +
\frac{1}{2}\int_{ I^2\setminus C_x} 1\, d\alpha\,
d\beta  \\ &=& \scalprod{y}{L_x y} + \frac{1}{2}\abs{I^2\setminus
C_x},\end{array}
\end{equation}
where Lemma \ref{lem:cons:scalprod_to_intsquare} was used to obtain
the last equability. For $y=x$, it follows from the definition of
$C_x$ that the above inequality is tight. In particular, the
following two relations hold for any $s$ and $t$:
\begin{equation*}
\begin{array}{lllll}
V(x_t)&=& \scalprod{x_t}{L_{x_t} x_t} &+&
\frac{1}{2}\abs{I^2\setminus
C_{x_t}}\\
V(x_s)&\leq &\scalprod{x_s}{L_{x_t} x_s} &+&
\frac{1}{2}\abs{I^2\setminus C_{x_t}}.
\end{array}
\end{equation*}
Taking $s=t+1$, we obtain
\begin{equation*}
\begin{array}{lll}
V(x_{t+1})-V(x_{t}) &\leq&  \scalprod{x_{t+1}}{L_{x_t} x_{t+1}} -
\scalprod{x_t}{L_{x_t} x_t} \\ &=&  2 \scalprod{\Delta x_t
}{\jnt{L_{x_t}} x_t}  + \scalprod{\Delta x_t}{L_{x_t}\Delta x_t},
\end{array}
\end{equation*}
where we have used the symmetry of $L_{x_t}$. It follows from
(\ref{eq:deltax=-dLx}) that $L_{x_t} x_t = -D_{x_{\jnt{t}}} \Delta
x_{\jnt{t}}$, so that
\begin{equation*}
\begin{array}{lll}
V(x_{t+1})-V(x_{t}) &\leq&   - 2 \scalprod{\Delta x_t }{D_{x_t}
x_t} + \scalprod{\Delta x_t}{L_{x_t}x_t} \\ &=&
\jnt{-}\scalprod{\Delta x_t}{(A_{x_t}+ D_{x_t}) \Delta x_t},
\end{array}
\end{equation*}
since $L_x = D_x- A_x$.
\end{IEEEproof}

As will be seen below, this result implies the convergence of
$\Delta x_t$ to 0 in a suitable topology. We now show that $L_xx$
is small only if $x$ is close to $F$, the set of functions taking
discrete values separated by at least 1. \modrev{\JNT{As a
corollary, we then obtain the} result that \jnt{$\bar F$} is
exactly the set of fixed points, as also shown in
\cite{Lorenz:2007}}. The intuition behind the proof of these
results \jnt{parallels our proof of
Theorem~\ref{thm:conv_discr-time_discr-agent},} and is as follows.
Consider an agent $\alpha$ with one of the smallest opinions
$x(\alpha)$. If the \jnt{change in $x(\alpha)$} is small, its
attraction by agents with larger opinions must be small, because
almost no agents have an opinion smaller than $x(\alpha)$.
Therefore, there must be very few agents with an opinion
significantly larger than $x(\alpha)$ that interact with $\alpha$,
while there might be many of them who have an opinion close to
$x(\alpha)$. In other words, possibly many agents have
approximately the same opinion $x(\alpha)$, and very few agents
have an opinion in the interval $[x(\alpha) + \epsilon,x(\alpha)
+1)$, \modif{so that $x$ is close to a function \JNT{in} $F$ in
that zone. Take now an agent $\alpha'$ with an opinion larger than
$x(\alpha)+1+\epsilon$, and such that very few agents have an
opinion in $(x(\alpha) +1 + \epsilon, x(\alpha'))$. This agent
interacts with very few agents having an opinion smaller than its
own.} Thus, if the change in such an agent's opinion is small,
this implies that its attraction by agents having larger opinions
is also small, and we can repeat the previous reasoning.

In order to provide a precise statement of the result, \jnt{we
associate an opinion function $x$ with a measure that describes
the distribution of opinions, and use a measure-theoretic
formalism.} For a \jnt{measurable} function $x:I \rightarrow
[0,L]$ (i.e., $x\in X_L$), and a \jnt{measurable} set $S\subseteq
[0,L]$, we let $\mu_x(S)$ be the Lebesgue measure of the set
$\{\alpha:x(\alpha) \in S\}$. By convention, we let $\mu(S)=0$ if
\modrev{$S\subseteq \Re\setminus[0,L]$}. To avoid confusion with
$\mu$, we use $\abs{S}$ to denote the standard Lebesgue measure of
a set $S$. We also introduce a suitable topology on the set of
opinion functions. We \jt{write} $x\leq_\mu \epsilon$ if
$|\{\alpha:x(\alpha)>\epsilon\}|\leq \epsilon$. Similarly,
\jt{$x<_\mu \epsilon$ if $|\{\alpha:x(\alpha) \geq \epsilon\}|<
\epsilon$, and} $x=_\mu 0 $ if $|\{\alpha:x(\alpha)\not  = 0\}| =
0$. We define the ``ball'' $B_\mu(x,\epsilon)$  as the set
$\{\jnt{y \in X_L}: |x-y| <_\mu \epsilon\}$. This allows us to
define a corresponding notion of limit. We say that $x_t
\rightarrow_\mu y$ if for all $\epsilon >0$, there is a $t'$ such
that for all $t>t'$ we have $x_t \in B_\mu(y,\epsilon)$. We write
$x_t\rightarrow_\mu S$ for a set $S$ if for all $\epsilon>0$,
there is a $t'$ such that for all $t>t'$, there is a $y\in S$ for
which $x_t\in B_\mu(y,\epsilon)$.

The result below, proved in Appendix \ref{appen:proof_decay_Lx},
states that the distance between $x\in X_L$ and $F$ (the subset of
$X_L$ consisting of functions taking discrete values separated by
at least 1) decreases to 0 (in a certain uniform sense) when $L_x
x\rightarrow_\mu 0$.

\begin{thm}\label{thm:cons:decay_Lx}
For any $\epsilon>0$, there exists a $\delta >0$ such that if
$|L_x x| <_\mu \delta$, then there exists some $s\in F$ with
$|x-s| <_\mu \epsilon$. In particular, if $L_x x =_\mu 0$, then
$x\in \bar F$.
\end{thm}
The next theorem compiles our convergence results.

\begin{thm}\label{thm:partial_conv_discr_time}
Let $(x_t)$ be a sequence of functions in $X_L$ evolving according
to the model (\ref{eq:def_cont_agent_discr_time}), and let $F$ be
the set of functions taking discrete values separated by at least
1. Then $(x_{t+1}-x_t) \rightarrow_\mu 0$ and
$\jnt{x_t\rightarrow_\mu F}$. \jnt{(In particular, periodic
trajectories, other than fixed points, are not possible.)}
Furthermore, $x$ is a fixed point of
(\ref{eq:def_cont_agent_discr_time}) if and only if $x\in \bar F$.
\end{thm}
\begin{IEEEproof}
We begin by proving the convergence of $\Delta x_t$. Suppose that
$\Delta x_t = (x_{t+1}-x_t) \rightarrow_\mu 0$ does not hold.
Then, there is an $\epsilon>0$ such that for arbitrarily large
$t$, there is a set of measure at least $\epsilon$ \JNT{such that
$\abs{\Delta x_t \jnt{(\alpha)}}>\epsilon$ for every $\alpha$ in
that set.} Consider such a time $t$. Without loss of generality,
assume that there is a set $S\subseteq I$ of measure at least
$\epsilon/2$ on which $\Delta x_t\jnt{(\alpha)}>\epsilon$.
(Otherwise, we can use a similar argument for the  set on which
$\Delta x_t \jnt{(\alpha)}< -\epsilon$.) \jnt{Fix some $L'>L$.}
For $i \in \{1,\dots, 2\ceil{ \jnt{L'}}\}$, let $A_i\subset I$ be
the set on which $x_t\in [(i-1)/2,i/2]$. For any $i$ and for any
$\alpha,\beta\in A_i$, there holds $\abs{x_t(\alpha)-x_t(\beta)}
<1$ and thus $(\alpha,\beta)\in C_{x_t}$. Therefore, $A_i^2
\subseteq C_{x_t}$ for all $i$. Moreover, the sets $A_i$ cover
$[0,1]$, so that $\sum_{i=1}^{2\ceil{L'}}|A_i\cap S|\geq |S| \geq
\epsilon/2$. Thus, there exists some $i^*$ such that
$|A_{\jnt{i^*}}\cap S| \geq \epsilon/(4\ceil{L'})$. We then have
\begin{equation*}\begin{array}{ll}
& \int_{ C_{x_t}}\prt{\Delta x_t(\alpha) + \Delta
x_t(\beta)}^2d\alpha\, d\beta \\ \geq& \int_{ (A_{i^*}\cap S)^2
}\prt{\Delta x_t(\alpha)
+ \Delta x_t(\beta)}^2d\alpha\, d\beta\\
\geq& 4\epsilon^2 |A_{i^*}\cap S|^2 \geq
{\epsilon^4}/{4\ceil{L'}^2}.\end{array}
\end{equation*}
Thus, if $\Delta x_t \rightarrow_\mu 0$ does not hold, then
$\int_{(\alpha,\beta)\in C_{x_t}}\prt{\Delta x_t(\alpha) + \Delta
x_t(\beta)}^2$ does not decay to 0, which contradicts Theorem
\ref{thm:lyapunov_discr_time}. \jnt{We conclude that $\Delta x_t
\rightarrow_\mu 0$. Using also \eqref{eq:deltax=-dLx} and the fact
$d_{x_t}(\alpha)\leq 1$, we obtain $L_{x_t}x_t \rightarrow_\mu 0$.
Theorem \ref{thm:cons:decay_Lx} then implies that $x_t
\rightarrow_\mu F$.}

\jnt{If $x\in \bar F$, it is immediate that $x$ is a fixed point.
Conversely, if $x_0=x$ is a fixed point, then $x_t=x_0$, a.e., for
all $t$. Then, the fact $x_t \rightarrow_\mu F$ implies that $x\in
\bar F$.}
\end{IEEEproof}

We note that the fact $x_t \rightarrow_\mu F$ means that the
measure $\mu_x$ associated with any limit point $x$ of $x_t$ is a
discrete measure whose support consists of values separated by at
least 1. Furthermore, it can be shown that at least one such limit
point exists, because of the semi-compactness of the set of
measures under the weak topology.

Theorem \ref{thm:partial_conv_discr_time} states that $x_t$ tends
to the set $F$, but \jt{does not guarantee convergence to an
element} of this set. We make the following conjecture, which is
currently unresolved.

\begin{conj}
Let $(x_t)$ be a sequence of functions in $X_L$, evolving according
to the model (\ref{eq:def_cont_agent_discr_time}). Then, there is a
function $x^*\in F$ such that $x_t\rightarrow_\mu x^*$.
\end{conj}

\subsection{Inter-cluster distances and stability of equilibria}

We have found that $x$ is a fixed point of
(\ref{eq:def_cont_agent_discr_time}) if and only if {it belongs to
$\bar F$}, that is, with the exception of a zero-measure set, the
range of $x$ is a discrete set of values that are separated by at
least one. As before, we will refer to these discrete values as
clusters. In this section, we consider the stability of
equilibria, and show that a condition on the inter-cluster
distances similar to the one in Theorem
\ref{thm:stab_disc_time_disc_agent} is necessary for stability.
Furthermore, we show that under a certain smoothness assumption,
the system cannot converge to a fixed point that does not satisfy
this condition.

In contrast to the discrete case, we can study the
continuous-agent model using the classical definition of
stability. We say that $s\in F$ is \modrev{\emph{stable}} if for
any $\epsilon
>0$, there is a $\delta>0$ such that for any $x_0\in
B_\mu(s,\delta)$, we have $x_t\in B_\mu(\jnt{s},\epsilon)$ for all
$t$. \modrev{It can be shown that} this notion encompasses the
stability with respect to the addition of a perturbing agent used
in Section \ref{sec:stab_discr_agents}. More precisely, if we view
the discrete-agent system as a special case of the continuum
model, stability under the current definition implies stability
with respect to the notion used in Section
\ref{sec:stab_discr_agents}. \modrev{The introduction of a
perturbing agent with opinion $\tilde x_0$ can indeed be simulated
by taking $x_0(\alpha)=s(\alpha)$ everywhere except on an
appropriate set of measure less than $\delta$, and $x_0(\alpha) =
\tilde x_0$ on this set.}  (However, the converse implication
turns out to not hold in some pathological cases. Indeed,
\jt{consider} two agents separated by exactly 2. They are stable
with respect to the definition of Section
\ref{sec:stab_discr_agents}, but not under the current definition.
This is because if we introduce a small measure set of additional
agents that are uniformly spread between the two original agents,
we will obtain convergence to a single cluster.) Moreover, it can
be verified that the notion of stability used here is equivalent
to both ${\cal L}_1$ and ${\cal L}_2$ stability. In the sequel,
and to simplify the presentation, we will neglect any zero measure
sets on which $\Delta x_t(\alpha) \not = 0$, and will give the
proof for a fixed point in $F$. The extension to fixed points in
$\bar F$ is straightforward. The proof of the following result is
similar to that of its discrete counterpart, the necessary part of
Theorem \ref{thm:stab_disc_time_disc_agent}, and is presented in
the Appendix \ref{appen:proof_stab_discr_cont_argent}.

\begin{thm}\label{thm:stab_discr_cont_agent}
Let $s\in F$ be a fixed point of
(\ref{eq:def_cont_agent_discr_time}), and let $a,b$ two values
taken by $s$. If $s$ is stable, then
\begin{equation}\label{eq:stab_condition}
\abs{b-a} \geq 1 +
\frac{\min\prt{\mu_s(a),\mu_s(b)}}{\max\prt{\mu_s(a),\mu_s(b)}}.
\end{equation}
\end{thm}

\providecommand{\regu}{regular}%

\jnt{With a little extra work, focused on the case where the
distance $|a-b|$ between the two clusters is exactly equal to 2,
we can show that the strict inequality version of condition
(\ref{eq:stab_condition}) is necessary for stability. We
conjecture that this strict inequality version is also
sufficient.}

\jnt{We will now proceed to show that under an additional
smoothness assumption \modrev{on the initial \JNT{opinion}
function}, we can never have convergence to a fixed point that
violates condition (\ref{eq:stab_condition}).} We start by
introducing the notion of a \regu{} opinion function. We say that
a function $x\in X_L$ is \emph{\regu{}} if there exist $M \geq m>
0$ such that any interval $J\subseteq [\inf_\alpha x, \sup_\alpha
x]$ satisfies $m\abs{J}\leq \mu_x(J) \leq M\abs{J}$. Intuitively,
a function is \regu{} if the set of opinions is connected, and if
the density of agents on any interval of opinions is bounded from
above and from below by positive constants. (In particular, no
single value is taken by a positive measure set of agents.) For
example, any piecewise differentiable $x\in X_L$ with positive
upper and lower bounds on its derivative is \regu.

We will show that if $x_0$ is \regu{} and if ($x_t$) converges,
then $x_t$ converges to an equilibrium satisfying the condition
(\ref{eq:stab_condition}) on the minimal distance between
opinions, provided that $\sup_\alpha x_t -\inf_\alpha x_t$ remains
always larger than 2. For convenience, we introduce  a nonlinear
update operator $U$ on $X_L$, defined by  $U(x) = x-D^{-1}_xL_xx =
D^{-1}_xA_xx$, so that the recurrence
(\ref{eq:def_cont_agent_discr_time}) can be written as $x_{t+1}
=U(x_t)$. The proof of the following proposition is presented in
Appendix \ref{appen:proof_regu_preserved}.

\begin{prop}\label{prop:regu_preserved}
Let $x\in X_L$ be a \regu{} function such that $\sup_\alpha x -
\inf_\alpha x > 2$. Then $U(x)$ is \regu.
\end{prop}

\jnt{We note that the assumption $\sup_\alpha x - \inf_\alpha x >
2$ in Proposition \ref{prop:regu_preserved} is necessary for the
result to hold. Indeed, if the opinion values are confined to a
set $[a,b]$, with $b-a=2-\delta<2$, then all agents  with opinions
in the set $[a+1-\delta,a+1]$ are connected with every other
agent, and their next opinions will be the same, resulting in a
non-\regu{} opinion function.}

\jnt{As a consequence of Proposition \ref{prop:regu_preserved},
together with Theorem \ref{thm:partial_conv_discr_time}, if $x_0$
is \regu, then there are two main possibilities: (i) There exists
some \jt{time} $t$ at which $\sup_\alpha x_t - \inf_\alpha x_t<2$.
In this case, the measure $\mu_{x_t}$ will have point masses
shortly thereafter, and will eventually converge to the set of
fixed points with at most two clusters. (ii) Alternatively, in the
``regular'' case, we have $\sup_\alpha x_t - \inf_\alpha x_t
\jt{>} 2$ for all times. Then, every $x_t$ is \regu, and
convergence cannot take place in finite time. Furthermore, as we
now proceed to show, convergence to a fixed point that violates
the stability condition (\ref{eq:stab_condition}) is impossible.
Let us note however that tight conditions for a sequence of
\regu{} functions to maintain the property $\sup_\alpha x_t -
\inf_\alpha x_t \jt{>} 2$ at all times appear to be difficult to
obtain.}

\begin{thm}\label{thm:no_conv_2_unstab_cont}
Let $(x_t)$ be a sequence of functions in $X_L$ that evolve according
to (\ref{eq:def_cont_agent_discr_time}). We assume that $x_0$ is \regu{}
and that $\sup_\alpha x_t - \inf_\alpha x_t > 2$ for all $t$. If
$(x_t)$ converges, then it converges to a function $s \in F$ such
that
\begin{equation*}
\abs{b-a} \geq 1 +
\frac{\min\prt{\mu_s(a),\mu_s(b)}}{\max\prt{\mu_s(a),\mu_s(b)}},
\end{equation*}
for any two \jnt{distinct} values  $a$, $b$, with  $\mu_s(a),\mu_s(b)>0$.
In particular, if $\mu_s(a) =\mu_s(b)$, then $\abs{b-a}\geq 2$.
\end{thm}
\begin{IEEEproof}
Suppose that $(x_t)$ converges to some $s$. It follows from
Theorem \ref{thm:partial_conv_discr_time} that $s\in F$, and from
Proposition \ref{prop:regu_preserved} that all $x_t$ are \regu.
Suppose now that $s$ violates the condition in the theorem, for
some $a$, $b$, with $a<b$. \jt{Then,} $b-a <2$, and we must have
$\mu_s\prt{(a,b)} = 0$ because all discrete values taken by $s$
(with positive measure) must differ by at least 1. We claim that
there exists a positive length interval $J\subseteq (a,b)$ such
that $\mu_{x_{t+1}}(J) \geq \mu_{x_{t}}(J)$ whenever $x_t\in
B_\mu(s,\epsilon)$, for a sufficiently small $\epsilon>0$. Since
$x_t$ converges to $s$, this will imply that there exists a finite
time $t^*$ after which $\mu_{x_{t}}(J)$ is nondecreasing, \jnt{and
$\liminf_{t\to\infty} \mu_{x_{t}}(J) \geq \mu_{x_{t^*}}(J)>0$. On
the other hand, since $\mu_s((a,b))=0$, $\mu_{x_{t}}(J)$ must
converge to zero. This is a contradiction and establishes the
desired result.}

\jnt{We now establish the above claim.} Let
$c=\frac{\mu_s(a)a+\mu_s(b)b}{\mu_s(a)+\mu_s(b)}$ be the weighted
average of $a$ and $b$. \jnt{The fact that the condition in the
theorem is violated implies (cf.\ \eqref{eq:condition}) that
$|c-a|<1$ and $|c-b|<1$.} Let $\delta>0$ be such that $c-\delta+1
> b$ and $c+\delta-1 < a$, and consider the interval
$J=[c-\delta,c+\delta]$.  For any $x\in B_\mu(s,\epsilon)$, we
have
\begin{eqnarray*}
\mu_x([a-\epsilon,a+\epsilon])\in [\mu_s(a) - \epsilon,\mu_s(a) + \epsilon],\\
\mu_x([b-\epsilon,b+\epsilon])\in [\mu_s(b) - \epsilon,\mu_s(b) + \epsilon],\\
\mu_x\prt{(a-1,b+1)\setminus \prt{[a-\epsilon, a+\epsilon]\cup
[b-\epsilon,b+\epsilon]} }\leq \epsilon,
\end{eqnarray*}
where we have used the fact that the values taken by $s$ are
separated by at least 1. \jnt{Suppose now that $\epsilon$ is
sufficiently small so that $c-\delta+1 > b+\epsilon$ and
$c+\delta-1 < a-\epsilon$. This implies that for every $\gamma$
such that $x(\gamma) \in J$, we have $(a-\epsilon,
b+\epsilon)\subseteq (x(\gamma)-1,x(\gamma)+1)$. If $\epsilon$
were equal to zero, we would have $u_x(d)=c$. When $\epsilon$ is
small, the location of the masses at $a$ and $b$ moves by an
$O(\epsilon)$ amount, and an additional $O(\epsilon)$ mass is
introduced. The overall effect is easily shown to be $O(\epsilon)$
(the detailed calculation can be found in
\cite{Hendrickx:2008phdthesis}). Thus, $|(U(x))(\gamma)-c|$ is of
order $O(\epsilon)$. When $\epsilon$ is chosen sufficiently small,
we obtain $c-\delta \leq (U(x))(\gamma) \leq c+\delta$, i.e.,
$(U(x))(\gamma) \in J$ for all $\gamma$ such that $x(\gamma) \in
J$. This implies that $\mu_{U(x)}(J)\geq \mu_x(J)$, and completes
the proof.}
\end{IEEEproof}

\section{{Relation between the discrete and the continuous-agent models}}\label{sec:discr-time_link}

We now analyze the extent to which the continuous-agent model
(\ref{eq:def_cont_agent_discr_time}) can be viewed as a limiting
case of the discrete-agent model
(\ref{eq:def_discrete-time_system}), when the number of agents
tends to infinity. As already explained in Section
\ref{sec:discr-time_cont_agent}, the continuous-agent model can
simulate exactly the discrete-agent model. \jnt{In this section,
we are interested in the converse; namely, the extent to which a
discrete-agent model can describe, with arbitrarily good
precision, the continuous-agent model.} We will rely on the
following result on the continuity of the update operator.

\begin{prop}\label{prop:continuity_at_regu}
Let $x\in X_L$ be a \regu{} function. Then, the update operator
$U$ is continuous at $x$ with respect to the norm
$\norm{\cdot}_\infty$. More precisely, for any $\epsilon >0$ there
exists some $\delta>0$ such that if $\norm{y-x}_\infty \leq \delta
$ then $\norm{U(y)-U(x)}_\infty \leq \epsilon$.
\end{prop}
\begin{IEEEproof}
Consider a \regu{} function $x\in X_L$, and an arbitrary
$\epsilon>0$. Let $\delta$ be smaller than ${m\epsilon}/{25M}$,
where $m$ and $M$ (with $m\leq M$) are the bounds in the
definition of \regu{} opinion functions applied to $x$. We will
show that if a function $y\in X_L$ satisfies $\norm{x-y}_\infty
\leq \delta$, then $\norm{U(y)-U(x)}_\infty\leq\epsilon$.

Fix some $\alpha \in I$, and let $S_{x},S_y\subseteq I$ be the set
of agents connected to $\alpha$ according to the interconnection
topologies $C_x$ and $C_y$ defined by $x$ and $y$, respectively.
We let $S_{xy} = S_{x}\cap S_y$, $S_{x\setminus y} = S_x \setminus
S_{xy}$ and $S_{y\setminus x} = S_y \setminus S_{xy}$. Since
$\norm{x-y}_\infty \leq \delta$, the values
$\abs{x(\alpha)-x(\beta)}$ and $\abs{y(\alpha)-y(\beta)}$ differ
by at most $2\delta$, for any $\beta\in I$. As a consequence, if
$\beta \in S_y$, then $ \abs{x(\alpha)-x(\beta)} \leq
\abs{y(\alpha)-y(\beta)} + 2 \delta$. Similarly, if $\beta \not
\in S_y$, then $ \abs{x(\alpha)-x(\beta)} \geq
\abs{y(\alpha)-y(\beta)} - 2 \delta$. Combining these two
inequalities with the definitions of $S_{xy}$, $S_{x\setminus y}$,
and $S_{y\setminus x}$, we obtain
\begin{small}
\begin{equation*}
\begin{array}{l}
[x(\alpha) - 1 + 2\delta,x(\alpha) + 1 - 2\delta ] \subseteq
x(S_{xy}) \subseteq [x(\alpha) - 1 ,x(\alpha) + 1 ],\\
x(S_{x\setminus y} ) \subseteq [x(\alpha)-1,
x(\alpha)-1+2\delta]\cup [x(\alpha)+1-2\delta, x(\alpha)+1],\\
x(S_{y\setminus x}) \subseteq [x(\alpha)-1-2\delta,
x(\alpha)-1]\cup [x(\alpha)+1, x(\alpha)+1+2\delta].
\end{array}
\end{equation*}\end{small}
\modifnj{Since $x$ is \regu, we have $|S_{xy}| \geq m(2-4\delta)
\geq m$ and $\abs{S_{x\setminus y}},\abs{S_{y\setminus x}}\leq
M4\delta$.} Let now $\bar x_{xy}$ and $\bar x_{x\setminus y}$ be
the average value of $x$ on $S_{xy}$ and $S_{x\setminus y}$,
respectively. Similarly, let $\bar y_{xy}$, and $\bar
y_{y\setminus x}$ be the average value of $y$ on $S_{xy}$ and
$S_{y\setminus x}$. Since $\norm{x-y}_\infty \leq \delta$, $\bar
x_{xy}$ and $\bar y_{xy}$ differ by at most $\delta$. It follows
from the definition of the model
(\ref{eq:def_cont_agent_discr_time}) that
\begin{equation*}
\begin{array}{lll}
(U(x))(\alpha) &=& \bar x_{xy} +\frac{\abs{S_{x\setminus
y}}}{|S_{xy}|+ \abs{S_{x\setminus y}}} (\bar x_{x\setminus y}-\bar
x_{xy}),\\
(U(y))(\alpha)& =& \bar y_{xy} + \frac{\abs{S_{y\setminus
x}}}{|S_{xy}|+ \abs{S_{y\setminus x}}} (\bar y_{y\setminus x}-\bar
y_{xy}).\end{array}
\end{equation*}

\jnt{It can be seen that} $\abs{\bar x_{x\setminus y}-\bar
x_{xy}}\leq 3$ and $\abs{\bar y_{y\setminus x}-\bar y_{xy}}\leq
3$, from which we obtain that $\abs{(U(y))(\alpha)-
(U(x))(\alpha)}$ is upper
\begin{equation*}
\abs{\bar x_{xy} - \bar y_{xy}} + 3\frac{\abs{S_{y\setminus
x}}}{|S_{xy}|} + 3\frac{\abs{S_{x\setminus y}}}{|S_{xy}|} \leq
\delta + 6\frac{4M\delta}{m} \leq \epsilon.
\end{equation*}
where we have used the fact that $\abs{\bar x_{xy}-\bar
y_{xy}}\leq \delta$. Since \jnt{the above} is true for any
$\alpha\in I$, we conclude that $\norm{\jnt{U(y)-U(x)}}_\infty
\leq \epsilon$.
\end{IEEEproof}

Let $U^t:X_L\to X_L$ be the composition of the update operator,
defined by $U^t(x)=U\prt{U^{t-1}(x)}$, so that $U^t(x_0) = x_t$.
Proposition \ref{prop:continuity_at_regu} is readily extended to a
continuity result for $U^t$.

\begin{cor}\label{cor:reg}
Let $x_0\in X_L$ be a \regu{} function such that $\sup_\alpha U^t(x)
- \inf_\alpha U^t(x) >2$ for every $t\geq 0$. Then for any finite
$t$, $U^t$ is continuous at $x$ with respect to the
 norm $\norm{\cdot}_\infty$.
\end{cor}
\begin{IEEEproof}
Since $x$ is \regu{} and since $\sup_\alpha U^t(x) - \inf_\alpha
U^t(x)>2$ for all $t$, Proposition \ref{prop:regu_preserved}
implies that all $U^t(x)$ are \regu. Proposition
\ref{prop:continuity_at_regu} then implies that for all $t$, $U$
is continuous at $U^t(x)$, and therefore the composition $U^t$ is
continuous at $x$.
\end{IEEEproof}

Corollary \ref{cor:reg} allows us to prove that, in the regular
case, and for any given finite time horizon, the continuous-agent
model is the limit of the discrete-agent model, as the number of
agents grows. \jnt{To this effect, for any given partition of
$I=[0,1]$ into $n$ disjoint sets $J_1,\ldots,J_n$, we define an
operator ${\cal G}: \Re^n\to X$ that translates the opinions in an
$n$-agent system to an opinion function in the continuous-agent
model. More precisely, for a vector $\hat x\in\Re^n$ and any
$\alpha\in J_i$, we let $({\cal G} \hat x)(\alpha)$ be equal to
the $i$th component of $\hat x$.}

\begin{thm}\label{thm:discr_appr_cont}
Let $\jnt{x_0}\in X_L$ be a \regu{} function and assume that
$\sup_\alpha x_t - \inf_\alpha x_t >2$ \jnt{for $t\leq t^*$.}
Then, the sequence $\prt{x_t}$, $t=1,\ldots,t^*$, can be
approximated arbitrarily well  by a sequence $\prt{\hat x_t}$ of
opinion vectors evolving according to
(\ref{eq:def_discrete-time_system}), in the following sense. For
any $\epsilon >0$, there \jnt{exists some $n$, a partition of $I$
into $n$ disjoint sets $J_1,\ldots,J_n$, and a vector $\hat x_0\in
[0,L]^n$ such that the sequence of vectors $\hat x_t$ generated by
the discrete-agent model (\ref{eq:def_discrete-time_system}),
starting from $\hat x_0$, satisfies $\norm{x_t - {\cal G} \hat x_t
}_{\infty}\leq \epsilon$, for $t=1,\ldots,t^*$.}
\end{thm}
\begin{IEEEproof}
Fix $\epsilon>0$. Since all $U^t$ are continuous at $x_0$,  there
is some $\delta>0$ such that if $\norm{y-x_0}_{\infty}\leq
\delta$, then $\norm{U^t(y)-x_t}_\infty \leq \epsilon$, for $t\leq
t^*$. Since $x_0$ is \regu, we can divide $I$ into subsets
$J_1,J_2,\dots,J_n$,  so that $|J_i| = 1/n$ for all $i$, and
$|x_0(\alpha)-x_0(\beta)|\leq \delta$ for all $\alpha$, $\beta$ in
the same set $J_i$. (This is done by letting $c_i$ be such that
$\mu_{x_0}([0,c_i])=i/n$, and defining $J_i=\{\alpha: c_{i-1} \leq
x_0(\alpha) \leq c_i\}$, where $n$ is sufficiently large.) We
define $\hat x_0 \in [0,L]^n$ by letting its $i$th component be
equal to $c_i$. We then have $\norm{x_0- {\cal G}\hat
x_0}_{\infty}\leq\delta$. This implies that $\norm{x_t-U^t({\cal
G}\hat x_0)}_{\infty}\leq \epsilon$, for $t\leq t^*$. Since the
continuous-agent model, initialized with a discrete distribution,
simulates the discrete-agent model, we have $U^t({\cal G}\hat
x_0)={\cal G}\hat x_t$, and  the desired result follows.
\end{IEEEproof}

Theorem \ref{thm:discr_appr_cont} supports the intuition that for
large values of $n$, the continuous-agent model behaves
approximatively as the discrete-agent model, over any finite
horizon. In view of Theorem \ref{thm:stab_discr_cont_agent}, this
suggests that the discrete-agent system should always converge to
a stable equilibrium (in the sense defined in Section
\ref{sec:discr_time_discr_agents}) when $n$ is sufficiently large,
as stated in Conjecture \ref{conj:stab_discr_ag_discr_time}, and
observed in many examples (see, e.g., Figure
\ref{fig:unstab2stab_n_increases_disc_time}). Indeed, Theorem
\ref{thm:stab_discr_cont_agent} states that under the regularity
assumption, the continuum system cannot  converge to an
equilibrium that does not satisfy condition
(\ref{eq:stab_condition}) on the inter-cluster distances.
\jnt{However, this argument does not translate to a proof \jt{of
the conjecture} because the approximation property in Theorem
\ref{thm:discr_appr_cont} only holds over a finite time horizon,
and does not necessarily provide information on the limiting
behavior.}

\section{Conclusions and Open questions}\label{sec:concl}

We have analyzed the model of opinion dynamics
(\ref{eq:def_discrete-time_system}) introduced by Krause,
\jnt{from several angles.} Our motivation was to provide an
analysis of a simple multi-agent system with an endogenously
changing interconnection topology while taking explicitly
advantage of the topology dynamics, \jnt{something that is rarely
done in the related literature.}

We focused our attention on an intriguing phenomenon, the fact
that equilibrium inter-cluster distances are usually significantly
larger than 1, and typically close to 2. We proposed an
explanation of this phenomenon based on a notion of stability with
respect to the addition of a perturbing agent. We showed that such
stability translates to a certain lower bound on the inter-cluster
distances, with the bound equal to 2 when the clusters have
identical weights. We also discussed the conjecture that when the
number of agents is sufficiently large, the system converges to a
stable equilibrium for ``most'' initial conditions.

To avoid granularity problems linked with the presence or absence
of an agent in a particular region, we introduced a new opinion
dynamics model that allows for a continuum of agents. For this
model we proved that under some regularity assumptions, there is
always a finite density of agents between any two clusters during
the convergence process. As a result, we could prove that such
systems never converge to an unstable equilibrium. We also proved
that the continuous-agent model is indeed the limit of a discrete
model, over any given finite time horizon, as the number of agents
grows to infinity. These results provide some additional support
for the conjectured, but not yet established, generic convergence
to stable equilibria.

We originally introduced the continuous-agent model as a tool for
the study of the discrete-agent model, but it is also of
independent interest and raises some challenging open questions.
\modrev{An important one is the question of whether the
continuous-agent model is always guaranteed to converge.} (We only
succeeded in establishing convergence to the set of fixed points,
not to a single fixed point.)

\jt{Finally, the study of the continuous-agent model suggests some
broader questions. In the same way that the convergence of the
discrete-agent model can be viewed as a special case of
convergence of inhomogeneous products of stochastic matrices, it
may be fruitful to view the convergence of the continuous-agent
model as a special case of convergence of inhomogeneous
compositions of stochastic operators, and to develop results for
the latter problem. }

The model (\ref{eq:def_discrete-time_system}) can of course be
extended to higher dimensional spaces, as is often done in the
opinion dynamics literature (see \cite{Lorenz:2007} for a survey).
Numerical experiments again show the emergence of clusters that
are separated by distances significantly larger than 1. The notion
of stability with respect to the addition of an agent can also be
extended to higher dimensions. However, stability conditions
become more complicated, and in particular cannot be expressed as
a conjunction of independent conditions, one for each pair of
clusters. For example, it turns out that adding a cluster to an
unstable equilibrium may render it stable
\cite{Hendrickx:2008phdthesis}. In addition, a formal analysis
appears difficult because \jnt{in $\Re^n$, with $n>1$, the support
of the opinion distribution can be connected without being convex,
and convexity is not necessarily preserved by our systems. For
this reason, even under ``regularity'' assumptions,} the presence
of perturbing agents between clusters is not guaranteed.

\begin{IEEEbiography}
[
{\includegraphics[width=1in,height=1.25in,clip,keepaspectratio]{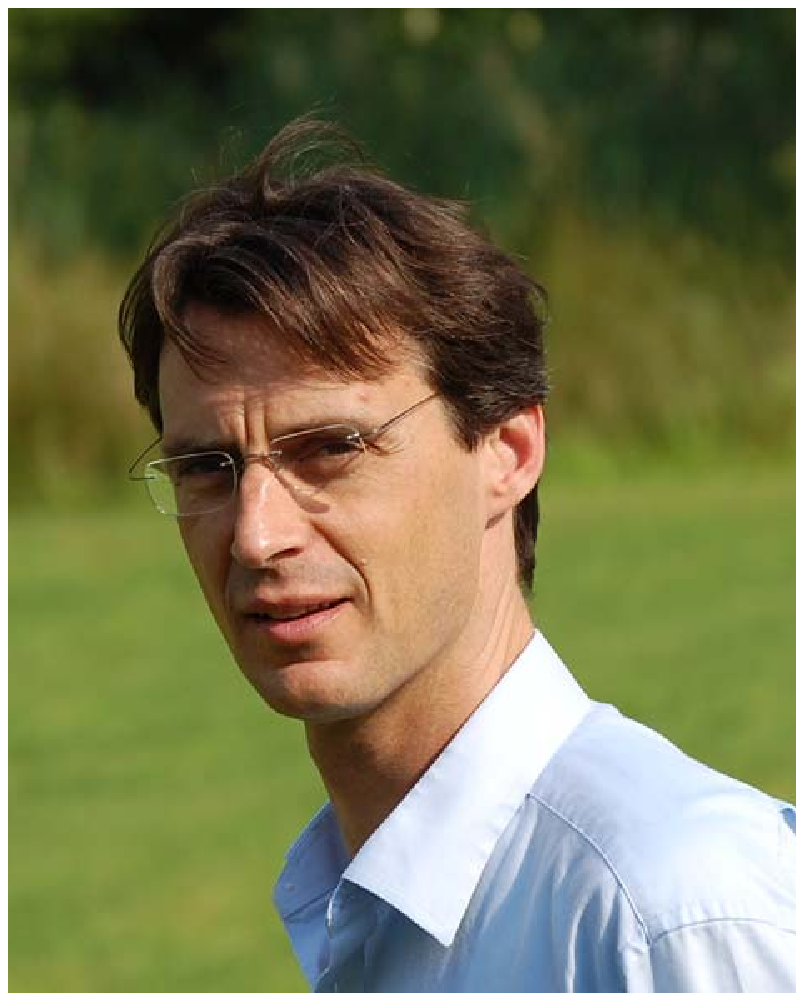}}
] {Vincent D. Blondel} received the M.Sc. degree in mathematics
from Imperial College, London, U.K., in 1990 and the Ph.D. degree
in applied mathematics from the Université catholique de Louvain,
Louvain-la-Neuve, Belgium, in 1992.

He was a Visiting Researcher at the Royal Institute of Technology,
Stockholm, Sweden, and at the Institut National de Recherche en
Informatique et en Automatique (INRIA), Rocquencourt, France.
During 2005-2006, he was an Invited Professor and a Fulbright
Scholar at Massachusetts Institute of Technology, Cambridge. He is
currently a Professor and Department Head at the Université
catholique de Louvain, Louvain-la-Neuve, Belgium.

Dr. Blondel was the recipient of the Prize Wetrems of the Belgian
Royal Academy of Science, the Society for Industrial and Applied
Mathematics (SIAM) Prize on Control and Systems Theory, and the
Ruberti Prize in Systems and Control of the IEEE in 2006.
\end{IEEEbiography}

\begin{IEEEbiography}
[
{\includegraphics[width=1in,height=1.25in,clip,keepaspectratio]{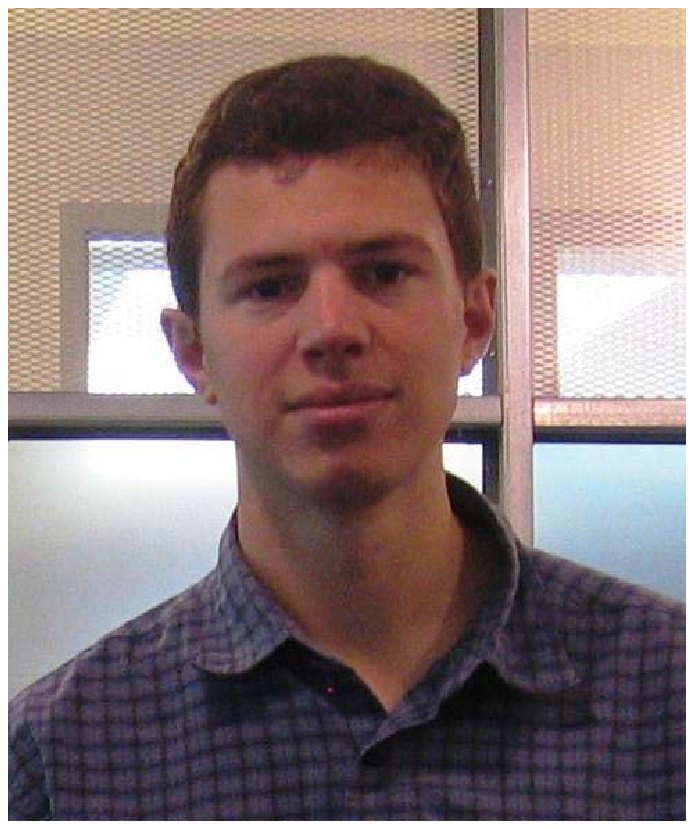}}
] {Julien M. Hendrickx}

\modrev{received an engineering degree in applied mathematics and
a PhD in mathematical engineering from the Université catholique
de Louvain, Belgium, in 2004 and 2008, respectively.

He has been a visiting researcher at the University of Illinois at
Urbana Champaign in 2003-2004, at the National ICT Australia in
2005 and 2006, and at the Massachusetts Institute of Technology in
2006 and 2008. He is currently a postdoctoral fellow at the
Laboratory for Information and Decision systems of the
Massachusetts Institute of Technology, and holds postdoctoral
fellowships of the F.R.S.-FNRS (Fund for Scientific Research) and
of Belgian American Education Foundation.

Doctor  Hendrickx was the recipient of the 2008 EECI award for the
best PhD thesis in Europe in the field of Embedded and Networked
Control. }
\end{IEEEbiography}

\begin{biography}[
{\includegraphics[width=1in,height=1.25in,clip,keepaspectratio]{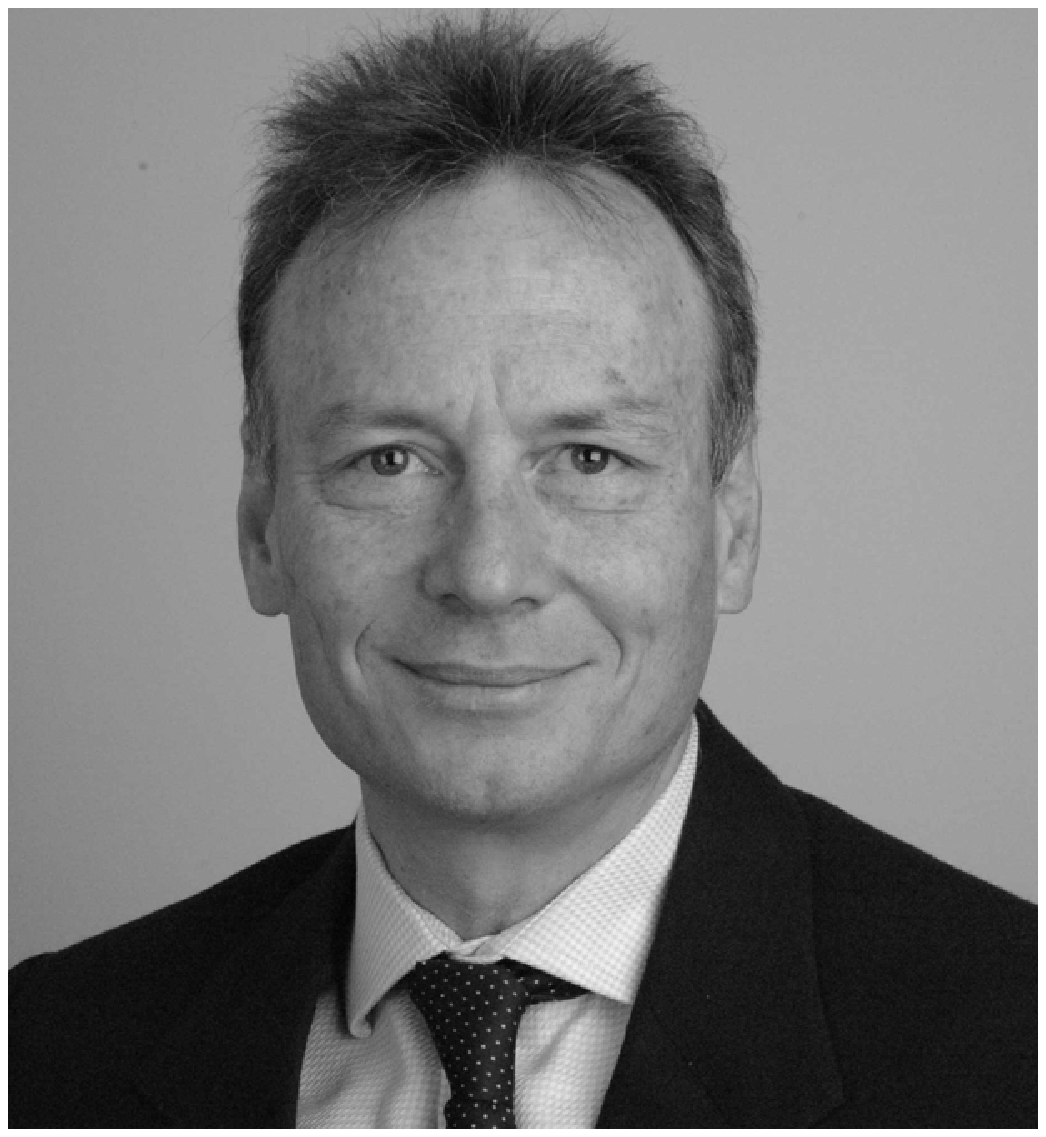}}
]{John N. Tsitsiklis} (F'99) received the B.S. degree in
mathematics and the B.S., M.S., and Ph.D. degrees in electrical
engineering from the Massachusetts Institute of Technology (MIT),
Cambridge, in 1980, 1980, 1981, and 1984, respectively. He is
currently a Clarence J.~Lebel Professor with the Department of
Electrical Engineering, MIT. He has served as a Codirector of the
MIT Operations Research Center from 2002 to 2005, and in the
National Council on Research and Technology in Greece (2005-2007).
His research interests are in systems, optimization,
communications, control, and operations research. He has
coauthored four books and more than a hundred journal papers in
these areas.

Prof.~Tsitsiklis was a recipient of an Outstanding Paper Award
from the IEEE Control Systems Society (1986), the M.I.T. Edgerton
Faculty Achievement Award (1989), the Bodossakis Foundation Prize
(1995), and the INFORMS/CSTS Prize (1997). He is a member of the
National Academy of Engineering. Finally, in 2008, he was
conferred the title of Doctor honoris causa, from the Universit\'e
catholique de Louvain
\end{biography}

 \newpage
 \appendix
\subsection{Proof of Theorem \ref{thm:cons:decay_Lx}}
\label{appen:proof_decay_Lx}

Before proceeding to the main part of the proof, we start with an
elementary lemma.

\providecommand{\lf}[2]{\hat L_{#1}\prt{#2}}%
\providecommand{\lmu}[1]{\lf{\mu}{#1}}%
\providecommand{\lmup}[1]{\hat L^+_{\mu}\prt{#1}}%
\providecommand{\lmum}[1]{\hat L^-_{\mu}\prt{#1}}%
\providecommand{\DI}[1]{\Delta_{#1}}%

\begin{lem}\label{lem:cons:boundsKDelta}
For any real numbers $\epsilon>0$, $M>1$, and any \jnt{positive}
integer $N$, there
exists a $\DI{1}>0$ and a sequence $K_1,K_2,\dots K_N,$ such that:\\
a) $K_i >M$, for \jnt{$i=1,\ldots,N$;}\\
b) the sequence $(\DI{i})$ defined by $\DI{i+1} = 3 K_i \DI{i} +
\frac{1}{K_i}$ satisfies $\DI{i}K_i < \epsilon$, for
\jnt{$i=1,\ldots,N$;}
\end{lem}
\begin{IEEEproof}
We use induction. The result is obviously valid for $N=1$. We now
assume that it holds for some $N$, and prove that it also holds
for $N+1$. Choose some $K_{N+1}$ such that $K_{N+1} > M$. Using
the induction hypothesis, choose $\DI{1}$ and a sequence
$K_1,\dots, K_N$ so that for $i=1,\dots N$, we have $K_i \DI{i} <
\frac{\epsilon}{6 K_{N+1}}$ and $K_i >
\max\prt{\frac{2K_{N+1}}{\epsilon},M}$. The conditions on $K_i$
are satisfied for $i=1,\dots,N+1$, and so are those on $K_i
\DI{i}$ for $i=1,\dots,N$. The result then follows from

\begin{equation*}
\begin{array}{lll}

K_{N+1}\DI{N+1} &=& K_{N+1} \prt{3 K_N \DI{N} + \frac{1}{K_N}}
\\&<& K_{N+1} \prt{3 \frac{\epsilon}{6 K_{N+1}} +
\frac{\epsilon}{2K_{N+1}}} =\epsilon. \end{array}
\end{equation*}
\end{IEEEproof}

To simplify the presentation of the proof, we introduce some new
notation. For a given measure $\mu$, we define the function $\hat
L_\mu$ by
\begin{equation*}
\lf{\mu}{y} = \int_{\jnt{(y-1,y+1)}} (y-z)\, d\mu(z).
\end{equation*}
Thus, for any $\alpha \in I$, we have $(L_xx)(\alpha) =
\lf{\mu_x}{x(\alpha)}$, \modif{where $\mu_x$ is the measure
associated to $x$, defined by letting $\mu_x(S)$ be the Lebesgue
measure of the set $\{\alpha:x(\alpha) \in S\}$ for any measurable
set $S$.} \modifnj{Since no ambiguity is possible here as we only
use one such measure, we will refer to $\mu_x$ as $\mu$ in the
sequel.} We also define the \jnt{nonnegative} functions
\begin{equation*}
\hat L^+_{\mu}\prt{y} = \int_{(y,y+1)}(z-y)\, d\mu(z)\geq 0,
\end{equation*}
and
\begin{equation*}
\hat L^-_{\mu}\prt{y} = \int_{(y-1,y)}(y-z)\, d\mu(z)\geq 0,
\end{equation*}
so that $\hat L_\mu = \hat L_\mu^+ - \hat L_\mu^-$. \jnt{Using the
definition of the relation $<_\mu$, we} observe that if
$|L_xx|<_\mu \delta$, then the set \jnt{$$S =\Big\{ y\in [0,L] :
\abs{\lmup{y}-\lmum{y}} \geq \delta\Big\}$$ satisfies} $\mu(S) <
\delta$. As a consequence, \modif{ if $|L_xx|<_\mu \delta$,} then
for any $z\in [0,L]$
\jnt{at least one of the following must be true:} \\
(i) there exists some \jnt{$y\in [z,L]$} such that $\lmup{y} <
\lmum{y} + \delta$ and
$\mu\prt{ \jnt{[z,y} )}\leq \delta$; or,\\
(ii) we have $\mu{ \jnt{([}z,L])}<\delta$. \jnt{We will make use
of this observation repeatedly in the proof of Theorem
\ref{thm:cons:decay_Lx}, which is given below.}

\begin{IEEEproof}
\jnt{Without loss of generality, we assume that $L$ is integer.
(This is because the model with $L$ not integer can be viewed as a
special case of a model in which opinions are distributed on
$[0,\jt{\ceil{L}}]$.)} We fix some $\epsilon>0$, \jnt{and without
loss of generality we assume that $\epsilon<1/2$.} Using Lemma
\ref{lem:cons:boundsKDelta}, we form two sequences
$K_1,\dots,K_{{L}\jnt{+1}}$ and $\DI{1},\dots,\DI{{L}\jnt{+1}}$
that satisfy: (i) \jnt{$\DI{i+1} = 3 K_i \DI{i} + \frac{1}{K_i}$,
for $i=1,\ldots{L}$;} (ii) $K_i>({L}+1)/\epsilon$ and $K_i\DI{i}<
\epsilon$, for $i=1,\ldots{L}\jnt{+1}$. In particular,
$\DI{i}<\epsilon^2/({L}+1)$. We then choose some $\delta$ smaller
than  $\DI{i}/3$, for all $i$.  \jnt{We will prove the following
claim.} If $|L_xx| <_\mu \delta$, then there exists \jnt{some
$N\leq L+1$}, and two \jnt{nonde}creasing \jnt{finite} sequences,
\jnt{$(x_i)$ and $(y_i)$,} that satisfy \jnt{$$-1=y_0< 0\leq
x_1\leq y_1\leq \cdots \leq x_N \leq y_N,$$ the termination
condition $\mu\prt{(y_{\jnt{N}},L]}< \epsilon^2/({L}+1)$, and the
following additional conditions, for $i=1,\ldots,N$:}
\\\\
(a)   $\lmup{x_i}<\DI{i}$; \\ %
(b)   $x_i \geq y_{i-1} +1$;  \\%
(c)   $\mu\prt{[y_{i-1},x_i)}\leq \DI{i}-\delta$;\\%
(d)   $0\leq y_i-x_i\leq K_i\DI{i}< \epsilon$.\\\\
\jnt{The above claim, once established,} implies that $\mu$ is
``close'' to a discrete measure \jnt{whose support consists of}
values that are separated by at least 1, \jnt{and provides a proof
of the theorem.} \jnt{To see this, note that the length of} each
interval $[x_i,y_i]$ \jnt{is less than} $\epsilon$. Furthermore,
the  set $[0,L]\setminus \bigcup\jnt{_{i=1}^N}[x_i,y_i]$ \jnt{is
covered by disjoint intervals, of the form $[0,x_i)$,
$(y_{i-1},x_i)$, or $(y_N,L]$. Since intervals $(y_{i-1},x_i)$
have at least unit length, the overall number of \jt{such}
intervals is at most $L+1$. For intervals of the form $[0,x_i)$,
$(y_{i-1},x_i)$, condition (c) implies that their measure is
bounded above by $\DI{i}-\delta<\DI{i}<\epsilon^2/({L}+1)$. Recall
also the termination condition
$\mu\prt{(y_{\jnt{N}},L]}<\epsilon^2/({L}+1)$. It follows that the
measure of the set $[0,L]\setminus \bigcup_i[x_i,y_i]$ is at most
$\epsilon^2$, hence smaller than $\epsilon$.} Let $s\in F$ be a
function which for every $\alpha$ takes a value $x_i$ which is
closest to $x(\alpha)$. Since $x$ can differ from all $x_i$ by
more than $\epsilon$ only on a set of measure \jnt{smaller than}
$\epsilon$, it follows that $|x-s|\jnt{<}_\mu \epsilon$. Finally,
if $L_xx =_\mu 0$, then $|L_xx| <_\mu \delta$ for all positive
$\delta$. As a consequence, the distance between $x$ and $F$ is
smaller than any positive $\epsilon$ and is thus 0. \modif{Because
$\bar F$ is the closure of $F$, it follows then that $x\in \bar
F$.} \jnt{Thus, it will suffice to provide a proof of the claim.}

%
%
%
%
\jnt{We will now use a recursive construction to prove the claim.}
We initialize the construction as follows. \jnt{Since $|L_xx|
<_{\mu} \delta$,} there exists some $x_1 \jnt{\geq 0}$ such that
$\mu\prt{[0,x_1 \jnt{)}}\leq \delta$ and $\lmup{x_1}\jnt{<}
\lmum{x_1} + \delta$. Since $y_{\jnt{0}} = -1$,  $x_1$ satisfies
condition (b). \jnt{Since $\delta<\Delta_1 /3$, we have
$\mu([y_0,x_1))=\mu([0,x_1))\leq \delta \leq 2\Delta_1/3
-\delta<\Delta_1-\delta$, and condition (c) is satisfied.}
Moreover,
\begin{equation*}
\begin{array}{lll} \lmum{x_1} &=& \int_{(x_1-1,x_1)}(x_1 - z)\, d\mu(z)
\\&\leq&\int_{(x_1-1,x_1)}\, d\mu(z) \\&\leq& \mu(\jnt{[}0,x_1)) \leq
\delta.\end{array}
\end{equation*}
\jnt{Thus, $\lmup{x_1}< \lmum{x_1} + \delta \leq
2\delta<\Delta_1$, and condition (a) is also satisfied.}

%
%
%

We now assume that we have chosen nonnegative $x_1,\dots,x_{i}$
and $y_1,\dots,y_{i-1}$, so that $x_1,\ldots,x_{i-1}$ satisfy the
four conditions (a)-(d), and $x_i$ satisfies conditions (a)-(c).
We will first show that we can choose  $y_i$ to satisfy condition
(d). \jnt{Then, if $\mu\prt{(y_{i},L]}<\epsilon^2/(L+1)$, we will
set $N=i$, and terminate the construction. Otherwise, we will show
that we can choose $x_{i+1}$ to satisfy conditions (a)-(c), and
continue similarly. Note that if $y_L$ has been thus constructed
and the process has not yet terminated, then the property
$y_{i+1}\geq x_i\geq y_i+1$ (from conditions (b) and (d)) implies
that $y_{L+1}\geq L$, so that $\mu\prt{(y_{L+1},L]}=0$, which
\modif{satisfies} the termination condition. This shows that
indeed $N\leq L+1$, as desired. Because all the required
conditions will be enforced, this construction will indeed verify
our claim.}

%
%
%
%
\jnt{The argument considers separately two different cases. For
the first case, we assume} that $\mu\prt{[x_i,x_i+1)}\leq \delta +
\frac{1}{K_i}$, which means that very few agents have opinions
between $x_i$ and $x_i+1$. The construction described below is
illustrated in Figure \ref{fig:cons:construc_density_Lxx}(a). We
let $y_i = x_i$, so that condition (d) is trivially satisfied by
$y_i$. \jnt{If $\mu\prt{(y_{i},L]}<\epsilon^2/(L+1)$, we let $i=N$
and terminate. Suppose therefore that this is not the case. Then,
$y_i<L$, and
$$\mu\prt{[y_{i},L+1]}=\mu\prt{[y_{i},L]}\geq \frac{\epsilon^2}{L+1}
>\Delta_{i+1}.$$
We then have
\begin{equation*}
\begin{array}{lll}
\mu\prt{[y_{i}+1,L+1]} &\jt{=}& \mu\prt{[y_{i},L+1]}
-\mu\prt{[y_{i},y_i+1)}\\&>&\Delta_{i+1}-\delta
-\frac{1}{K_i}>\delta,
\end{array}
\end{equation*}
where the last inequality follows from the recurrence $\DI{i+1} =
3 K_i \DI{i} + \frac{1}{K_i}$ (cf.\ Lemma
\ref{lem:cons:boundsKDelta}), and the fact that $2\delta <
\DI{i}<K_i \DI{i} $ for all $i$. In particular, we must have
$y_i+1\leq L$. } Using the assumption $|L_xx| <_\mu \delta$, and
an earlier observation, we can find some $x_{i+1} \jnt{\geq
y_i+1}$ such that $\lmup{x_{i+1}} \jt{<} \lmum{x_{i+1}} + \delta$
and $\mu{([y_{i}+1,x_{i+1}))} \jnt{\leq}\delta$. Condition (b)
then trivially holds for $x_{i+1}$. \modif{Remembering that
$y_i=x_i$,} we also have
\begin{equation*}
\begin{array}{lll}
\mu\prt{[y_{i},x_{i+1})} &=& \mu\prt{[x_i,x_i+1)}+
\mu\prt{[y_i+1,x_{i+1})} \\ & \leq & \delta + \frac{1}{K_i} +
\delta \leq \DI{i+1} - \delta,
\end{array}
\end{equation*}
where the last inequality follows, as before, from the recurrence
$\DI{i+1} = 3 K_i \DI{i} + \frac{1}{K_i}$, \modif{and from $\delta
\leq K_i\DI{i}$.} As a result, \jnt{$x_{i+1}$} satisfies condition
(c). To prove \jnt{that $x_{i+1}$ satisfies} condition (a),
observe that
\begin{equation*}
\begin{array}{lll}
 \lmum{x_{i+1}} &=& \int_{
(x_{i+1}-1,x_{i+1})}(x_{i+1}-z)\,d\mu(z) \\&\leq&
\mu\prt{[x_{i+1}-1,x_{i+1})}\\&\leq&  \mu\prt{[y_{i},x_{i+1})}.
\end{array}
\end{equation*}
where the last inequality follows from condition (b) for
$x_{i+1}$. Because $x_{i+1}$ satisfies condition (c), we have
$\lmum{x_{i+1}} \leq \DI{i+1} - \delta$.  Then, condition (a) for
$x_{i+1}$ follows from the fact that $x_{i+1}$ has been chosen so
that $\lmup{x_{i+1}} \jt{<} \lmum{x_{i+1}} + \delta$.

%
%
%
%

\begin{figure}

\centering
\begin{tabular}{c}
\includegraphics[scale=.35]{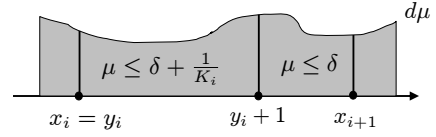}\\(a)\\\\
\includegraphics[scale=.35]{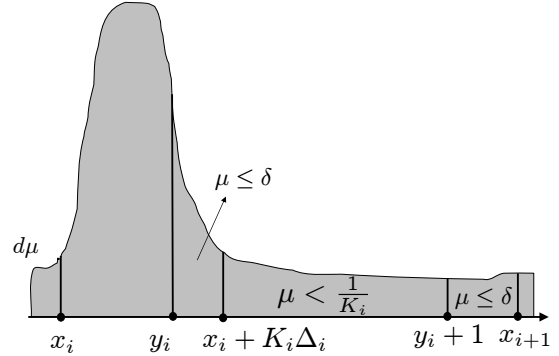}\\(b)
\end{tabular}

\caption[Representation of the iterative construction in the proof
of Theorem \ref{thm:cons:decay_Lx}.]{Illustration of the iterative
construction in the proof of Theorem \ref{thm:cons:decay_Lx}, when
(a) $\mu\prt{[x_i,x_i+1)}\leq \delta + \frac{1}{K_i}$, and (b)
$\mu\prt{[x_i,x_i+1)}> \delta + \frac{1}{K_i}$. For the sake of
clarity, the figure shows the density $d\mu$ as if it were
continuous, but the proof does not use the continuity, or even
existence, of a density.}\label{fig:cons:construc_density_Lxx}
\end{figure}

%
%
%
%

We now consider the second case, where
$\mu\prt{[x_i,x_i+1)}>\delta + \frac{1}{K_i}$. Our construction is
illustrated in Figure \ref{fig:cons:construc_density_Lxx}(b).
\jnt{We claim} that $\mu\prt{[x_i + K_i\DI{i}, x_i+1)} \leq
\frac{1}{K_i}$. This is because  otherwise we would have
\begin{equation*}
\begin{array}{lll}
\lmup{x_i} &=& \int_{(x_i,x_i+1)}(z-x_i)\, d\mu(z) \\ &\geq&
\int_{(x_i+K_i\DI{i},x_i+1)}(z-x_i)\, d\mu(z) \\ &\geq&
K_i\DI{i}\mu\prt{[x_i + K_i\DI{i}, x_i+1)} \\&>& \DI{i},
\end{array}
\end{equation*}
\jnt{contradicting condition (a) for $x_i$.} \jnt{This implies
that $\mu([x_i,x_i+K\Delta_i])>\delta$. It follows that we can
choose $y_i \geq x_i$} so that $y_i\leq x_i + K_i\DI{i}$,
$\mu\prt{[y_i,x_i + K_i\DI{i})}\leq \delta$, and $\lmup{y_{i}}\leq
\lmum{y_{i}} + \delta$. Then, condition (d) is satisfied by $y_i$.
\jnt{If $\mu\prt{[y_{i},L]}<\epsilon^2/(L+1)$, we let $i=N$ and
terminate. Suppose therefore that this is not the case. By the
same argument as for the previous case, we can} then choose
$x_{i+1}$ so that \jnt{$y_i +1 \leq x_{i+1}\leq L$,}
$\lmup{x_{i+1}}\leq \lmum{x_{i+1}} + \delta$, and
$\mu\prt{[y_{i}+1,x_{i+1})}<\delta$. Thus $x_{i+1}$ satisfies
condition (b).

To prove that $x_{i+1}$ satisfies the remaining two conditions,
(c) and (a), we need an upper bound on $\mu\prt{[x_{i}+1,y_i+1)}$.
Observe first that
\begin{equation}\label{eq:cons:lmup_yi_lower}
\begin{array}{llll}
\lmup{y_i} &=& \int_{ [y_i,y_i+1)} (z-y_i)\, d\mu(z) \\ &\geq&
\int_{[x_{i}+1,y_i+1) }(z-y_i)\, d\mu(z) \\ &\geq&
\prt{1+x_{i}-y_i}\mu\prt{[x_{i}+1,y_i+1)}\\&\geq&
\frac{1}{2}\mu\prt{[x_{i}+1,y_i+1)},
\end{array}
\end{equation}
where the last inequality \jnt{follows from the fact
$1+x_i-y_i\geq 1-\epsilon$} (condition (d)), and the fact that
$\epsilon$ was assumed  smaller than $\frac{1}{2}$. \jnt{Thus, to
derive upper bound on $\mu\prt{[x_{i}+1,y_i+1)}$ it will suffice
to derive an upper bound on $\lmup{y_i}$. We start with the
inequality, $\lmup{y_i} \leq \lmum{y_i}+\delta$, and also make use
of the fact $y_{i}-1 \geq x_i - 1 \geq y_{i\modif{-}1}$, which is
a consequence of conditions (b) and (d) for $x_i$.} We obtain
\begin{equation*}
\begin{array}{lll}
\lmum{y_i} &\leq& \int_{[y_{i-1},x_i)}(y_i-z)\, d\mu(z) +
\int_{[x_i,y_i)}(y_i-z)\, d\mu(z) \\&\leq &\mu\prt{[y_{i-1},x_i)}
+ \mu\prt{[x_i,y_i)} (y_i-x_i)
\\&\leq& \DI{i} - \delta + K_i\DI{i},
\end{array}
\end{equation*}
where the last inequality follows from conditions (c) and (d) for
$x_i$, and the fact that $\mu\prt{[x_i,y_i)} \leq
\mu\prt{[0,L]}=1$. Combining this with the lower bound
(\ref{eq:cons:lmup_yi_lower}) leads to
\begin{equation}\label{eq:cons:bound_mu_xi+1_yi+1}
\mu\prt{[x_{i}+1,y_i+1)} \leq 2(K_i+1) \DI{i},
\end{equation}
which is the desired upper bound on $\mu\prt{[x_{i}+1,y_i+1)}$.

We will now use the above upper bound to prove conditions (a) and
(c) for $x_{i+1}$. Observe that $\mu\prt{[y_i,x_{i+1})}$ can be
expressed as
\begin{equation*}
\begin{array}{ll}
 &\mu\prt{[y_i,x_i+K_i\DI{i})}+
\mu\prt{[x_i+K_i\DI{i},x_i+1)}\\ +& \mu\prt{[x_i+1,y_i+1)} +
\mu\prt{[y_i+1,x_{i+1})}.
\end{array}
\end{equation*}
Recall that $y_i$ has been chosen so that
$\mu\prt{[y_i,x_i+K_i\DI{i})}\leq \delta$, and $x_{i+1}$ so that
$\mu\prt{[y_i+1,x_{i+1})}\leq \delta$. Moreover,
$\mu\prt{[x_i+K_i\DI{i},x_i+1)}$ has been shown to be no greater
than ${1}/{K_i}$. It then follows from
(\ref{eq:cons:bound_mu_xi+1_yi+1}) that
\begin{equation*}
\mu\prt{[y_i,x_{i+1})} \leq 2\delta + \frac{1}{K_i}+ 2(K_i+1)
\DI{i}  \leq 3K_i\DI{i} + \frac{1}{K_i} - \delta,
\end{equation*}
where we have used the facts that $3\delta \leq \DI{i}$ and
$K_i\geq 3$. Condition (c) for $x_{i+1}$ follows from the property
$\DI{i+1} = 3 K_i \DI{i} + \frac{1}{K_i}$ in the definition of the
sequence $(\DI{i})$ (see Lemma \ref{lem:cons:boundsKDelta}). To
prove condition (a), \jnt{we observe} that
\begin{equation*}
\begin{array}{lll}
\lmum{x_{i+1}}&=&\int_{(x_{i+1}-1,x_{i+1})}(x_{i+1}-z)\,d\mu(z)
\\ &\leq& \mu\prt{(x_{i+1}-1,x_{i+1})},\end{array}
\end{equation*}
and then use conditions (b) and (c) for $x_{i+1}$ to obtain
\begin{equation*}
\lmum{x_{i+1}} \leq \mu\prt{[y_{i},x_{i+1})} \leq \DI{i}-\delta.
\end{equation*}
This completes the induction and the proof of Theorem
\ref{thm:cons:decay_Lx}.
\end{IEEEproof}

\subsection{\label{appen:proof_stab_discr_cont_argent}Proof of Theorem
\ref{thm:stab_discr_cont_agent}}

\begin{IEEEproof}
Suppose $s$ does not satisfy the condition of this theorem, and
that $a<b$. Since $s\in F$, we have $\mu_s\prt{(a,b)}=0$. Let
$S_a,S_b\subset I$ be two sets on which $s$ takes the values $a$
and $b$, respectively. We choose these sets so that the
\jnt{Lebesgue} measure of $S_a \cup S_b$ is $\delta$, and so that
the ratio \jnt{$|S_a|/|S_b|$} of their measures is equal to
${\mu_s(a)}/{\mu_s(b)}$. Let $x_0(\alpha)=s\jnt{(\alpha)}$ for
$\alpha\notin S_a\cup S_b$, and $x_0(\alpha) = \frac{\mu_s(a)a +
\mu_s(b)b}{\mu_s(a) + \mu_s(b)}$ for $\alpha\in S_a\cup S_b$.
Observe that $\frac{\mu_{x_0}(a)}{\mu_{x_0}(b)} =
\frac{\mu_s(a)}{\mu_s(b)}$.

As already discussed, when $x_0$ takes discrete values, the
evolution of $x_t$ is entirely characterized by the evolution of a
corresponding weighted discrete-agents system of the form
\eqref{eq:weighted_1D_model}. We can then apply the reasoning in
the proof of Theorem \ref{thm:stab_disc_time_disc_agent} to show
that the two clusters initially at $a$ and $b$ converge to a
single cluster. Since this can be done for any, arbitrarily small
$\delta >0$, $s$ is unstable.
\end{IEEEproof}

\subsection{\label{appen:proof_regu_preserved}Proof of Proposition
\ref{prop:regu_preserved}}

To ease the reading of the proof, we introduce a new notation. For
any $x\in X_L$, we let $u_x:[0,L]\to[0,L]$ be a function defined
so that $u_x(a)$ is the updated opinion of an agent that held
opinion $a$, namely
\begin{equation*}
u_x(a) = \frac{\jnt{\int_{a-1}^{a+1}}z\,
d\mu_{\jt{x}}\jt{(z)}}{\mu_x\prt{(a-1,a+1)}}.
\end{equation*}
As a consequence, $(U(x))(\alpha) = u_x(x(\alpha))$ and
$x_{t+1}(\alpha) = u_{x_t}\prt{x_t(\alpha)}$, for any $\alpha\in
I$.

\begin{IEEEproof}
Since $x$ is \regu, there exist $m$ and $M$, with $0<m\leq M$,
such that for any $[a,b]\subseteq [\inf_\alpha x,\sup_\alpha x]$
we have $m(b-a) \leq \mu_x([a,b]) \leq M (b-a)$. Let $\delta =
\min\{ \frac{1}{2}, \sup_{\alpha}x - \inf_{\alpha}x -2 \}$. We
first prove the existence of $M',m'> 0$ such that if $[a,b]
\subseteq [\inf_\alpha x,\sup_\alpha x]$ and $b-a< \delta$, then
$m'(b-a)\leq u_x(b) - u_x(a)\leq M'(b-a)$. \jnt{(The proof of the
upper bound amounts to noting that the numerator and denominator
in the definition of $u_x(a)$ are both Lipschitz continuous
functions of $a$, and that the denominator is bounded below by
$m$. The proof of the lower bound is essentially a strengthening
of the proof of Proposition \ref{prop:order_preservation}, which
only established that $u_x(b) - u_x(a)\geq 0$.)}

With our choice of $\delta$, we have either $a\geq \inf_\alpha x
+1$ or $b \leq \sup_\alpha x -1$. We only consider the second
case, so that $(a,b+1)\subseteq [\inf_\alpha x,\sup_\alpha x]$;
the first case can be treated similarly.  Let $\bar
\mu_{a\setminus
 b}= \mu_x\prt{(a-1,b-1]}$, $\bar \mu_{ab} = \mu_x\prt{(b-1,a+1)}$,  and $\bar \mu_{b\setminus a}=
\mu_x\prt{[a+1,b+1)}$. Let also $\bar x_{a\setminus b}$, $\bar
x_{ab}$, and $\bar x_{b\setminus a}$ be the \jnt{center of mass of
the opinions of those agents} whose opinions lie in the set
$(a-1,b-1]$, $(b-1,a+1)$, and $[a+1,b+1)$, respectively. (In case
$\bar \mu_{a\setminus b}=0$, we use the convention $x_{a\setminus
b}=b-1$.)

From the definition of $u_x$, we have
\begin{equation*}
u_x(a) = \frac{\bar\mu_{ab} \bar x_{ab} + \bar \mu_{a\setminus b}
\bar x_{a\setminus b} }{\bar\mu_{ab}+ \bar \mu_{a\setminus b} } =
\bar x_{ab} - \frac{ \bar \mu_{a\setminus b} (\bar x_{ab} - \bar
x_{a\setminus b} ) }{\bar\mu_{ab} + \bar \mu_{a\setminus b}},
\end{equation*}
and
\begin{equation*}
u_x(b) = \frac{\bar\mu_{ab} \bar x_{ab} + \bar \mu_{b\setminus a}
\bar x_{b\setminus a} }{\bar\mu_{ab}+ \bar \mu_{b\setminus a} } =
\bar x_{ab} + \frac{ \bar \mu_{b\setminus a} (\bar x_{b\setminus
a}- \bar x_{ab}) }{\bar\mu_{ab} + \bar \mu_{b\setminus a}}.
\end{equation*}
Note that $\jnt{a-1 \leq } \bar x_{a\setminus b} \leq \bar x_{ab}
\leq  \bar x_{b\setminus a} \jnt{\leq b+1}$, so that $\bar
x_{b\setminus a} - \bar x_{ab} \leq 2 + (b-a) \leq 3$, and
similarly, $\bar x_{ab}- \bar x_{a\setminus b} \leq  3$. From the
regularity assumption, we also have $\bar \mu_{b\setminus a}\leq
M(b-a)$ and $\bar \mu_{ab} = \mu_x\prt{(b-1,a+1)} \geq
\mu_x\prt{(a,a+1)} \geq m$. Thus,
\begin{equation*}
u_x(b)-u_x(a) \leq \jnt{3}\frac{ \bar \mu_{b\setminus a}
}{\bar\mu_{ab}} +\jnt{3} \frac{ \bar \mu_{a\setminus b}
}{\bar\mu_{ab}} \leq \frac{\jnt{3}M(b-a)}{m},
\end{equation*}
which proves the claimed upper bound with $M' = 3\frac{M}{m}$.

\jnt{For the lower bound, an elementary calculation shows that if
we have \jt{a} density function on the interval $[0,1]$, which
\jt{is} bounded above and below by $M$ and $m$, respectively, then
its center of mass is at least $m/2M$. By applying this fact to
the interval $(b-1,a+1)$ (which has length larger than 1), we
conclude that its center of mass, $\bar x_{ab}$ is at least $m/2M$
below the right end-point $a+1$. Since also $\bar x_{b\setminus
a}\geq a+1$, we have $\bar x_{b\setminus a}-\bar x_{ab} \geq
m/2M$, and
$$ u_x(b)\geq \bar x_{ab} + \frac{ \bar \mu_{b\setminus a} }
{\bar\mu_{ab} + \bar \mu_{b\setminus a}}\cdot
\frac{m}{2M} \geq u_x(a)+ \frac{ m(b-a)}{ \jt{3}M}\cdot
\frac{m}{2M}, $$ where the last inequality made use of the facts
$\mu_{b\setminus a}\geq m(b-a)$ and $\bar\mu_{ab} + \bar
\mu_{b\setminus a}\leq 3M$. This establishes the claimed lower
bound, with $m'=m^2/\jt{6}M^2$.}

\jnt{By splitting an interval $[a,b] \subseteq [ \inf_\alpha x,
\sup_\alpha x ]$ into subintervals of length bounded by $\delta$,
we see that the result $m'(b-a)\leq u_x(b) - u_x(a)\leq M'(b-a)$
also holds for general such intervals.} Consider now an interval
$[a',b'] \in [\inf_\alpha U(x),\sup_\alpha U(x)]$, and let $a =
\inf \{z\in [0,L]:u_x(z)\in [a',b']\}$ and $b =  \sup \{z\in
[0,L]:u_x(z)\in [a',b']\}$. As a consequence of the order
preservation property, $ u_x\prt{(a,b)}\subseteq [a',b']$, and
$[a',b']\subseteq [u_x(a),u_x(b)]$. \jnt{Since $x$ is \regu,} we
have  $\mu_x(a) = \mu_x(b) = 0$, which implies that
$\mu_{U(x)}([a',b']) = \mu_x([a,b]) \in [m(b-a),M(b-a)]$. Using
the bounds on $\frac{u_x(b)-u_x(a)}{b-a}$, we finally obtain the
desired result
\begin{equation*}
mm'(b'-a') \leq \mu_{U(x)}([a',b']) \leq MM' (b'-a').
\end{equation*}
\end{IEEEproof}

\end{document}